\documentclass[twocolumn,tighten]{aastex62}
\usepackage{natbib,amsmath}
\usepackage{epsfig}
\usepackage{color}
\usepackage{hyperref}

\graphicspath{{./Images/}}

\newcommand{\spitzer}{\textit{Spitzer}}

\newcommand{\herschel}{\textit{Herschel}}
\newcommand{\pabeta}{\mbox{\rm Pa$\beta$}}
\newcommand{\halpha}{\mbox{\rm H$\alpha$}}
\newcommand{\Ahalpha}{\mbox{\rm $A(H \alpha)$}}
\newcommand{\Apabeta}{\mbox{\rm $A(Pa \beta)$}}
\newcommand{\lunits}{\mbox{\rm erg~s$^{-1}$}}
\newcommand{\intunits}{\mbox{\rm erg~s$^{-1}$~cm$^{-2}$~sr$^{-1}$}}
\newcommand{\msun}{\mbox{\rm $M_{\odot}$}}

\newcommand{\code}[1]{\texttt{#1}}
\newcommand{\e}[1]{$\times\ 10^{#1}$}
\newcommand{\CASS}{\affiliation{Center for Astrophysics and Space Sciences, Department of Physics, University of California, San Diego\\9500 Gilman Drive, La Jolla, CA 92093, USA}}
\newcommand{\OSU}{\affiliation{Department of Astronomy, The Ohio State University\\4055 McPherson Laboratory, 140 West 18th Ave, Columbus, OH 43210, USA}}
\newcommand{\NRAO}{\affiliation{National Radio Astronomy Observatory, 520 Edgemont Road, Charlottesville, VA 22903, USA}}
\newcommand{\OAN}{\affiliation{Observatorio Astron\'{o}mico Nacional (IGN), C/ Alfonso XII 3, E-28014 Madrid, Spain}}
\newcommand{\MPIA}{\affiliation{Max Planck Institute for Astronomy, K\'onigstuhl 17, D-69117 Heidelberg, Germany}}

\newcommand{\Illinois}{\affiliation{Department of Astronomy, University of Illinois, Urbana, IL 61801, USA}}
\newcommand{\ALMA}{\affiliation{Joint ALMA Observatory, Alonso de C\'ordova 3107, Vitacura, Santiago, Chile}}

\shorttitle{\pabeta , \halpha ,  and Attenuation in NGC~5194 and NGC~6946}
\shortauthors{Kessler et. al}

\begin{document}

\title{\pabeta{}, \halpha{}, and Attenuation in NGC~5194 and NGC~6946}

\author{Sarah Kessler}
\OSU
\author{Adam Leroy}
\OSU
\author{Miguel Querejeta}
\OAN
\author{Eric Murphy}
\NRAO
\author{David Rebolledo}
\ALMA
\NRAO
\author{Karin Sandstrom} 
\CASS
\author{Eva Schinnerer}
\MPIA
\author{Tony Wong}
\Illinois

\begin{abstract}
We combine \textit{Hubble} Space Telescope (HST) Paschen $\beta$ (\pabeta{}) imaging with ground-based, previously published \halpha{} maps to estimate the attenuation affecting \halpha{}, \Ahalpha{}, across the nearby, face-on galaxies NGC~5194 and NGC~6946. We estimate \Ahalpha{} in $\sim$ 2,000 independent $2'' \sim 75$~pc diameter apertures in each galaxy, spanning out to a galactocentric radius of almost 10 kpc. In both galaxies, \Ahalpha{} drops with radius, with a bright, high attenuation inner region, though in detail the profiles differ between the two galaxies. Regions with the highest attenuation-corrected \halpha{} luminosity show the highest attenuation, but the observed \halpha{} luminosity of a region is not a good predictor of attenuation in our data. Consistent with much previous work, the IR-to-\halpha{} color does a good job of predicting \Ahalpha{}. We calculate the best-fit empirical coefficients for use combining \halpha{} with 8, 12, 24, 70, or 100$\mu$m to correct for attenuation. These agree well with previous work but we also measure significant scatter around each of these linear relations. The local atomic plus molecular gas column density, $N(H)$, also predicts \Ahalpha{} well. We show that a screen with magnitude $\sim 0.2$ times that expected for a Milky Way gas-to-dust value does a reasonable job of explaining \Ahalpha{} as a function of $N(H)$. This could be expected if only $\sim 40\%$ of gas and dust directly overlap regions of \halpha{} emission.
\end{abstract}



\section{Introduction}
\label{sec:intro}

The star formation rate (SFR) is of fundamental interest to many fields of astronomy. In order to properly understand star formation, chemical enrichment, feedback, and the evolution of star-forming galaxies, the SFR must be well measured and understood.

There are numerous methods used to derive the SFR \citep[for reviews see][]{KENNICUTT12,CALZETTI13}. The most direct approach is to count the number of young stars of known age and divide by the age since formation. Unfortunately, individual stars are difficult to resolve and characterize in other galaxies. A less direct, but more widely applicable, method is to trace star formation via the highly visible signatures of young, massive stars. When present, young stars contribute most of the ultraviolet (UV) light from a galaxy. This high-energy emission is heavily absorbed by foreground dust and reprocessed into the infrared (IR). As a result, both the UV and IR are frequently employed to estimate the SFR. However, dust attenuation and sensitivity to the recent star formation history can confuse the use of the UV, while IR emission can also trace emission from older stars. As a result of these ambiguities, tracers of the number of ionizing photons, including hydrogen recombination lines and free-free emission, are often considered the least ambiguous tracer of star formation.

Ionizing photons are mostly produced by massive stars (M \textgreater 10 \msun) which live only for a short amount of time. The ionizations caused by these photons are balanced by recombinations. The subsequent emission of hydrogen recombination lines creates an observable tracer of the most recent \citep[$\sim$ 10 Myr, e.g.,][]{KENNICUTT98} star formation.  Because the ionized gas is a direct result of short lived massive stars, recombinations lines are less affected by variation in the star-formation history than UV or IR emission. Disentangling ionizations due to AGN and shocks from those caused by massive stars does introduce some uncertainties, as does leakage from {\sc Hii} regions and emission from diffuse ionized gas (DIG). But in this paper we mainly focus on emission from bright regions in the disks of nearby star-forming galaxies, where AGN and LINER emission are not a major concern. Our aperture-based methodology will tend to minimize the contributions from DIG. 

The presence of dust complicates SFR estimation based on recombination lines. Dust absorbs much of the optical recombination line emission before it reaches the observer \citep[again see reviews by][]{KENNICUTT12,CALZETTI13}. Near-infrared (NIR) recombination lines are much less effected by dust attenuation than optical lines. For example, \pabeta{}, the NIR line that we use in this paper, suffers $\sim 3 \times$ less attenuation than H$\alpha$. By combining NIR and optical recombination lines, and making assumptions about the extinction curve and dust geometry, one can calculate the amount of foreground dust present. Therefore, the combination of NIR and optical recombination lines is considered a highly reliable measure of the SFR.

In this paper, following \citet{TURNER87},\citet{BECK84},\citet{HO90}, \citet{Scoville_2003}, \citet{CALZETTI05},  \citet{CALZETTI07}, \citet{KENNICUTT07}, and \citet{LI13}, among others, we adopt this approach. We combine optical \halpha{} and NIR \pabeta{} recombination lines to estimate the effects of dust on recombination line emission. \halpha{} is the $3 \rightarrow 2$ recombination line of hydrogen (at 6562.8 \AA). \pabeta{} is the $5 \rightarrow 3$ recombination line (at 1.282 \micron). Assuming case B recombination with a known temperature, the ratio of \halpha{} to \pabeta{} in the absence of dust is known theoretically \citep{HUMMER87}. Assuming an extinction curve \citep[this paper uses ][]{CCM89}, this known ratio of \halpha{} to \pabeta{} allows us to accurately calculate the attenuation affecting \halpha, \Ahalpha . 

A similar approach using the Balmer decrement, (\halpha/H$\beta$), has been widely used \citep[e.g.,][among many others]{BLANC09, KRECKEL13, CROXALL15,CATALAN-TORRECILLA15, TOMICIC19}. The Balmer decrement is less sensitive to dust attenuation than the \pabeta{}-to-\halpha{} ratio. 
For our adopted extinction curve and Case B recombination at $T=10,000$~K and density 1,000~cm$^{-3}$, \halpha{} is 1.3 times less attenuated by dust than H$\beta$ \citep{CCM89}. Because \pabeta{} is emitted at a longer wavelength, for the same conditions, \pabeta{} will be 3.2 times less attenuated by dust than \halpha{}.

Despite their power, galaxy-wide measurements of NIR recombination lines are rare. The NIR is hard to observe from the ground due to the atmosphere, hence the popularity of the Balmer decrement. The \textit{Hubble} Space Telescope's (HST) NICMOS instrument provided groundbreaking observations of the 4$\xrightarrow{}$3 recombination line of hydrogen, Pa$\alpha$ ($\lambda$ = 1.875\micron), but the instrument was limited in its field of view \citep[e.g.,][]{CALZETTI07,KENNICUTT07}. 

In this paper, we take advantage of HST's Wide Field Camera 3 (WFC3) to observe the \pabeta{} line using narrow band imaging techniques. We present new observations of NGC~6946 and analyze archival observations of NGC~5194. Both galaxies are prime targets for recombination line observations. These are two of the nearest ($d \sim 8$~Mpc), face-on, massive actively star-forming galaxies. Both have been used to calibrate SFR estimators in the past \citep[][]{CALZETTI05,CALZETTI07,KENNICUTT07,BLANC09,MURPHY11,LI13}. We combine the new \pabeta{} images with previously published \halpha{} to study attenuation in {\sc Hii} regions over the whole disk area of both galaxies.

Using these measurements, we estimate \Ahalpha{} across these two galaxies. We place apertures across the galaxy, covering all bright \halpha{}-emitting regions. We measure the fluxes of \pabeta{} and \halpha{} in each aperture and use these to calculate \Ahalpha. Then, we measure the distribution of \Ahalpha{} and its dependence on radius. We compare \Ahalpha{} to the local IR-to-\halpha{} color to test how well the local IR-to-\halpha{} predicts \Ahalpha{ } \citep[following][]{CALZETTI05,CALZETTI07,KENNICUTT07}. We also compare \Ahalpha{} to maps of the neutral gas column density. This allows us to evaluate how well gas can be used to predict attenuation and place constraints on the relative position of the neutral gas and {\sc Hii} regions.

This paper is organized as follows: Section \ref{sec:data} describes the new and archival data used in the analysis. Section \ref{sec:analysis} explains the how we place our apertures and estimate attenuation. Section \ref{sec:results} describes our results. 

\section{Data}
\label{sec:data}

\begin{figure*}
\centering
\includegraphics[width = 0.475\textwidth]{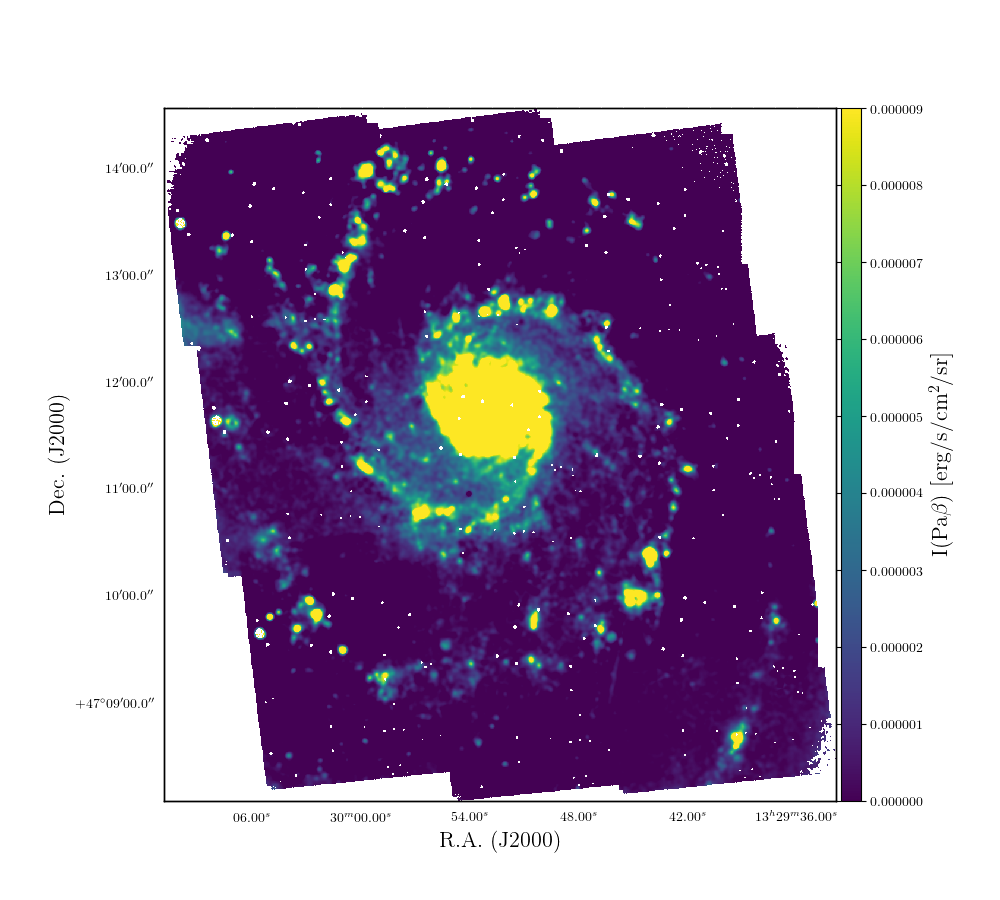}
\includegraphics[width = 0.475\textwidth]{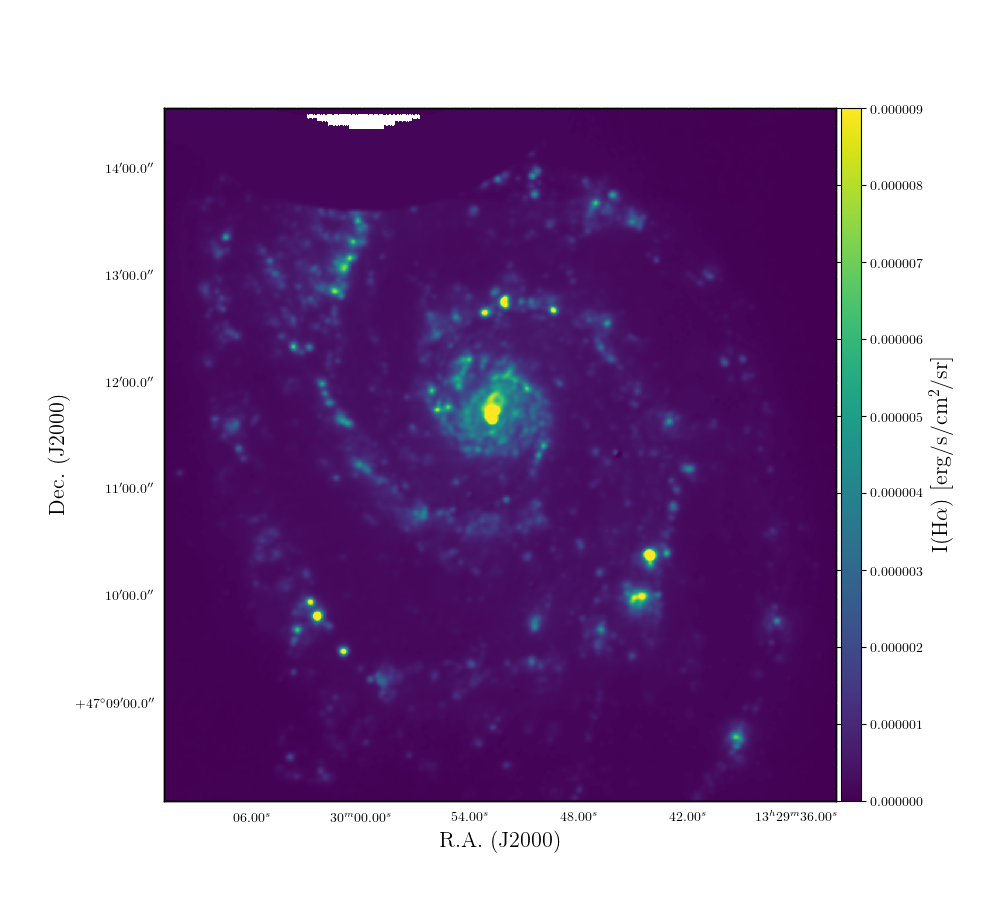}
\includegraphics[width=0.475\textwidth]{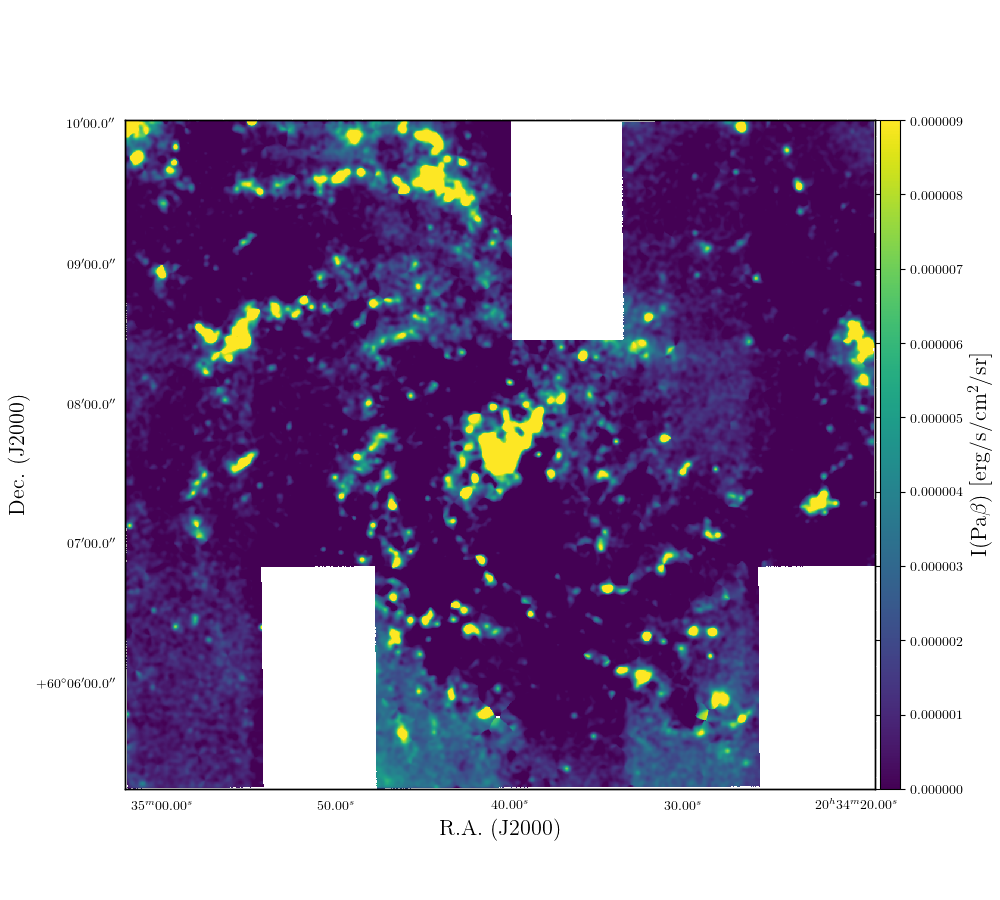}
\includegraphics[width=0.475\textwidth]{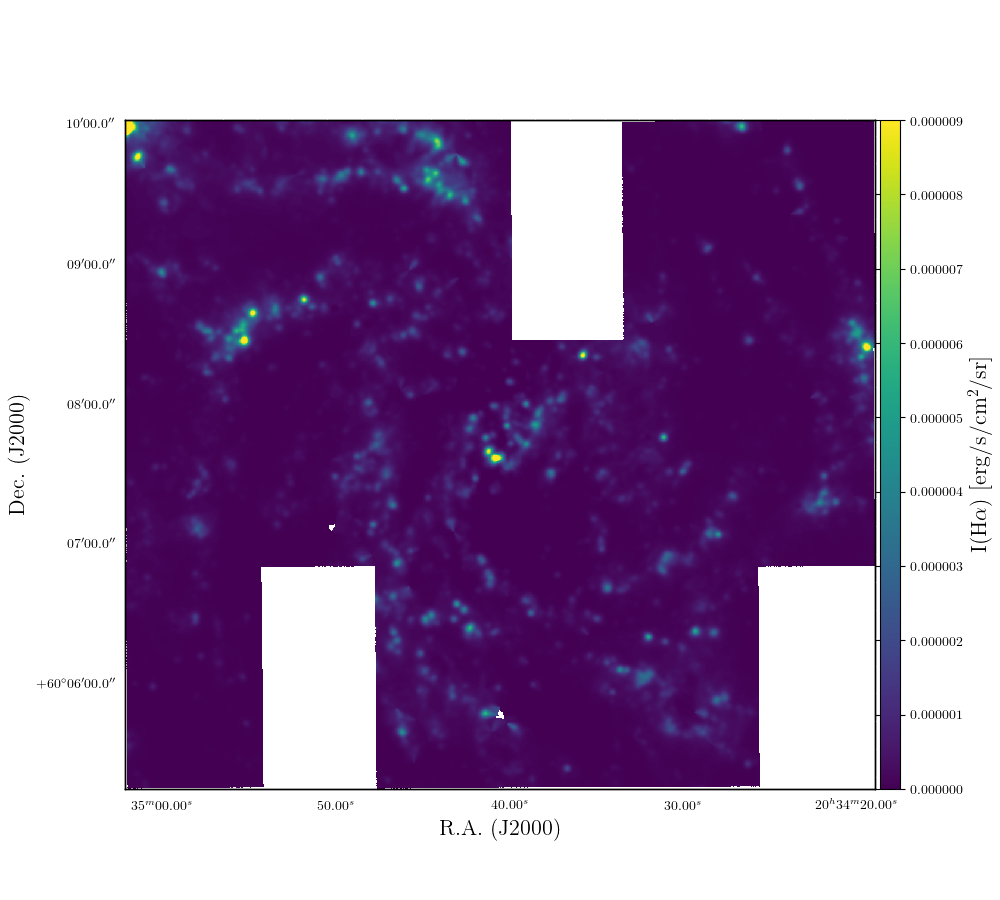}
\caption{\textbf{Images of NGC~5194 (top) and NGC~6946 (bottom) in \pabeta{} (left) and \halpha{} (right) at our working 2\arcsec{} resolution.} The images are scaled so that the intensity scales would match in the absence of attenuation. The blank regions in the NGC~6946 image reflect persistent background problems in the HST data (see Appendix).}
\label{fig:NGC5194 Mosaic}
\end{figure*}

\subsection{Targets}

NGC~5194 (M51) and NGC~6946 are two of the closest massive, face-on spiral galaxies. They are both gas-rich and dusty and are therefore natural targets to study the effects of attenuation on recombination line emission. Given this, both galaxies have been targets of many previous studies focused on ionizing photons. For example, NGC~5194 has targeted by \citet[][]{KENNICUTT07}, \citet{CALZETTI07}, \citet{BLANC09}, and \citet{QUEREJETA2019}. NGC~6946 has been studied by \citet{CALZETTI07}, \citet{MURPHY11}, \citet{LI13}, and \citet{LONG2019}, among others. Table \ref{tab:Adopted Properties} gives the adopted properties for each galaxy.

\begin{deluxetable}{lcc}
\tabletypesize{\scriptsize}
\tablecaption{Adopted Galaxy Properties \label{tab:Adopted Properties}}
\tablewidth{0pt}
\tablehead{
\colhead{Quantity} &
\colhead{NGC~5194} &
\colhead{NGC~6946}
}
\startdata
R.A. (J2000) & 13h 29m 52.7s & 20h 34m 52s \\
Dec. (J2000) & +47d 11h 43s & +60d 9m 11s\\
Dist. & 8.58$\pm$0.10~Mpc& 7.72$\pm$ 0.32~Mpc\\
P.A. & 173$^\circ$ & 243$^\circ$\\
Inclination & 21$^\circ$ & 33$^\circ$\\
Foreground $A_V$ & 0.096~mag & 0.938~mag 
\enddata
\tablecomments{Distances from \citet{MCQUINN16} for NGC~5194 and \citet{ELDRIDGE19} for NGC~6946. Orientations from \citet{COLOMBO14} for NGC~5914, \citet{DEBLOK08} for NGC~6946. Foreground attenuation from \citet{SCHLAFLY11}.}
\end{deluxetable}

\subsection{HST observations}

We observed NGC~6946 (Proposal ID 14156) in \pabeta{} with WFC3. NGC~5194 was observed in \pabeta{} by HST as part of Proposal 12490 (P.I.: J. Koda). We retrieve these data from the archive and analyze them in conjunction with our new observations of NGC~6946. 

Both observations used WFC3's F110W and F128N filters and have a native PSF of 0.1\arcsec. F110W is WFC3's wide J filter. It has a peak wavelength of 1.150 \micron{} and a FWHM of 0.500 \micron. We use this filter to trace the stellar continuum in both galaxies. We will refer to F110W as the OFF filter. F128N is WFC3's narrow \pabeta{} filter. It has a peak wavelength of 1.284 \micron{} and a FWHM of 0.0128 \micron. We will refer to F128N as the ON filter. The observed wavelength for \pabeta{} changes by $\leq$ 0.002 \micron\ relative to the rest value for the redshifts of NGC~5194 ($z = 0.00154 \pm 0.00001$) and NGC~6946 ($z = 0.0013 \pm 0.00001$) \citep{EPINAT08}. Therefore we expect the full line from both galaxies to lie within the F128N filter with little to no change in the transmission.

NGC~6946 was observed in nine fields which covered the area of the galaxy. NGC~5194 was observed in ten fields which covered both M51a and its companion, M51b. For our analysis we use only the nine fields associated with M51a. Using the OFF filter, each NGC~6946 field was observed for 456 seconds and each NGC~5194 field was observed for 612 seconds. In the ON filter, each NGC~6946 field was observed for 1059 seconds, and in NGC~5194 each field was observed for 2012 seconds. All observations employed a dithering pattern, and for each frame, associated sets of dithered images were combined using the WFC3 \code{astrodrizzle} pipeline.

The appendix steps through the details of our reduction and processing of the HST data, including background and continuum subtraction. The end result of this processing is a continuum subtracted \pabeta{} image for each galaxy with a $2''$ (FWHM) Gaussian PSF. The final \pabeta{} line images are shown in Fig. \ref{fig:NGC5194 Mosaic}.

\subsection{Archival \halpha\ data}
\label{sec:halpha data}

We combine archival narrow-band \halpha{} images with our \pabeta{} maps to estimate the amount of dust attenuation in each region. For NGC~6946, we utilize the \halpha{} map from \citet{LONG2019}. This map has already been flux calibrated and corrected for atmospheric transmission. Additionally, the filter width is small enough that significant {\sc Nii} contamination should not be an issue \citep[for more details see][]{LONG2019}. Based on fitting the stars in this image, we estimate the PSF to have FWHM 0.855\arcsec.

We perform the continuum subtraction ourselves. We use the narrow band OFF image described in \citet{LONG2019}. Because this filter does not overlap \halpha, there is no risk of oversubtracting the line. Using the same methods as in Section \ref{sec: continuum subtraction}, we derive and ON-to-OFF ratio of $\beta$ = 0.011 and use this to subtracting the stellar continuum. This produces an \halpha{} line image that looks properly subtracted when visually inspected.

At our working 2\arcsec{} (FWHM) resolution, the NGC~6946 map has 1$\sigma$ scatter of 2.35\e{-7}~\intunits . When we place 2\arcsec{} diameter apertures in apparently empty parts of the image, the rms scatter in luminosity is 8.65\e{34} \lunits .

For NGC~5194 we utilize the \halpha\ map from  the \textit{Spitzer} Infrared Nearby Galaxies Survey \citep[SINGS;][]{KENNICUTT03,KENNICUTT07}. This map has been corrected for {\sc Nii} contamination adopting a ratio {\sc Nii}/\halpha{} = 0.6 \citep{KENNICUTT07}, which introduces an error of $\sim$10\% or less for most regions. As a check, we verified that the SINGS map that we use matches the high quality spectral measurements from the \citet{BLANC09} IFU map in the region of overlap.

At 2\arcsec{} resolution the \halpha{} map for NGC~5194 has 1$\sigma$ scatter of 1.31\e{-7} \intunits . When we place 2\arcsec{} diameter apertures on apparently empty parts of the image, the rms scatter in the luminosity is 6.26\e{35} \lunits .

\subsection{Matching point spread functions}

We convolve the \pabeta{} and \halpha{} maps from their native resolutions to share a Gaussian PSF with FWHM 2\arcsec{}. For NGC~5194 we also make a version of the maps with a FWHM of 3\arcsec{}. By matching the PSFs at these resolutions, we can directly compare the much higher resolution HST data to the ground-based \halpha{} data and the VLA data. To carry out the convolution, we use astropy's \citep{ASTROPYV2} \code{convolve} and \code{Gaussian2DKernel} functions. At NGC~5194, $2'' \approx 83$~pc and $3'' \approx 125$~pc. At NGC~6946, $2'' \approx 73$~pc.

Despite our processing, the background in the \pabeta{} maps remains unstable at a low level on large scales. This is not an issue when studying bright emission using aperture photometry at 2-3\arcsec{} resolution. However, when we convolve the \pabeta{} images to much coarser resolution, uncertainties in the background dominate the image. These appear as large-scale gradients, most likely caused by a time variable background (TVB, see Section \ref{sec:NGC 6946 sky correction}). This prevents convolution of the \pabeta{} data to lower resolution to match the PSF of the IR images.

The backgrounds of the \halpha{} maps appear more stable. We convolve the \halpha{} data to 4, 7.5, 8, 9 and 11\arcsec{} resolution to match that of the IR data.

\subsection{Archival infrared data}

We compare the attenuation estimated from recombination lines to archival maps of IR emission at $8-100\micron$. Both galaxies were observed by \textit{Spitzer} at 8 and 24 \micron{} as part of SINGS \citep{KENNICUTT03}. NGC~6946 was observed at 70 and 100\micron by \herschel{} as part of KINGFISH \citep[NGC~6946;][]{KENNICUTT11}. NGC~5194 was also observed by \textit{Herschel} as part of the Very Nearby Galaxy Survey \citep[VNGS, NGC~5194;][]{BENDO12,MENTUCHCOOPER12} but only at 70$\mu$m. We also compare to 12\micron{} data from WISE, using the maps constructed by \citet{LEROY19}.

The PSFs of these IR data are very large compared to those of the \pabeta{} or \halpha{} maps. Using the kernels of \citet{ANIANO11}, the maps have all been convolved to have Gaussian beams with the following FWHM resolutions: $4''$ ($8$~\micron), $7.5''$ ($12\micron$), $11''$ (24\micron ), $8''$ (70~\micron), and $9''$ (100\micron ). As mentioned above, we also construct versions of the \halpha{} map at each of these resolutions.

\subsection{Archival 33~GHz data}

We compare our recombination-line based results to the 33~GHz radio continuum map of NGC~5194 by \citet{QUEREJETA2019}. In normal star-forming galaxies, radio emission at 33~GHz contains a large contribution from free-free emission. This free-free emission arises mostly from {\sc Hii} regions and is not attenuated by dust \citep[e.g.,][]{CONDON92, MURPHY11, MURPHY12}. As such, radio continuum emission provides a powerful alternative probe on the ionizing photon production rate. In combination with \halpha{}, it offers an alternative way to estimate \Ahalpha.

\citet{QUEREJETA2019} describe the observations and processing of these data. This is one of the widest area 33~GHz maps of any nearby galaxy and covers a substantial fraction of our \pabeta{} maps. The version of the map that we use has a 3\arcsec{} PSF and an rms noise of 6$\mu$Jy/beam. The observations are sensitive to spatial scales up to 44\arcsec, or 1.6~kpc. Given our aperture photometry-based approach (see Sec. \ref{sec:apertures}), we do not expect spatial filtering to represent a significant concern.

The 33~GHz map contains both free-free and synchrotron emission. For tracing ionizing photons, we are only interested in free-free emission, and so need to correct for contamination by synchrotron emission. We adopt a thermal fraction of 70\% for all apertures. This is simple, consistent with the free-free fraction found for 33~GHz peaks by \citet{QUEREJETA2019}. We show below that this leads to good agreement with the \pabeta\ map.

We do caution that this fixed free-free fraction might be too high for some apertures in the center of NGC~5194. There the central AGN may contribute more contaminating synchrotron emission. Linden et al. (in prep.) suggest that the thermal fraction might be $\sim 55$\% at 33~GHz in this region. We make no additional corrections, preferring a simple fixed 70\% free-free fraction as an easy-to-reproduce check on our measurements.

\subsection{Archival gas data}
\label{sec:column}

We explore how attenuation relates to the local gas column density, which we measure from 21-cm and CO emission line maps. We use 21-cm line maps from The {\sc Hi} Nearby Galaxy Survey \citep[THINGS][]{WALTER08} to trace the atomic hydrogen ({\sc Hi}) column density in both galaxies. We use the ``natural weighted'' maps, which have resolution of $11''.9 \times 10''.0$ for NGC~5194 and $6''.0 \times 5''.6$ for NGC~6946. Both resolutions are coarser than our working $2''$ resolution. We assume that the atomic gas is smooth below this resolution \citep[e.g., see][]{LEROY13B}. We convert these data to units of column density assuming optically thin {\sc Hi} and taking no account of helium. In the center of NGC~6946, \textbf{the 21-cm line} goes into absorption against the bright continuum associated with the nuclear starburst. Thus we do not derive an {\sc Hi} column density for the innermost point.

We use CO to estimate the column density of molecular hydrogen, H$_2$. For each galaxy, we use two maps, one at low resolution and one at higher resolution. The low resolution maps come from the HERACLES Survey \citep[][]{LEROY09}. For NGC~5194 this is a reprocessing of the data of \citet{SCHUSTER07}. HERACLES observed the CO~(2-1) line with resolution $13''.3$. High resolution CO~(1-0) maps also exist for both galaxies. PAWS \citet{SCHINNERER13}, observed CO~(1-0) emission from the inner part of NGC~5194 at $1'' \sim 40$~pc resolution. We use a version of this map convolved to $2''.2 \sim 80$~pc resolution. \citet{DONOVAN-MEYER12}, \citet{REBOLLEDO15}, and Rebolledo et al. 2020 (in prep) used CARMA  to observe CO~(1-0) from NGC~6946. We use a version of their map convolved to $5''.2 \sim 140$~pc resolution.

We convert from CO to H$_2$ column density using a standard ``Galactic''  conversion factor \citep{BOLATTO13}. For CO~(1-0) and not including any contribution from helium this is $\alpha_{\rm CO} = 3.2$~M$_\odot$~pc$^{-2}$~(K~km~s$^{-1}$)$^{-1}$. For the CO~(2-1) maps we assume a CO~(2-1)/(1-0) ratio of $0.7$ \citep[e.g., see][]{LEROY13A}, implying $\alpha_{\rm CO}^{2-1} = 4.6$~M$_\odot$~pc$^{-2}$~(K~km~s$^{-1}$)$^{-1}$. After converting to mass surface density, we re-express the maps equivalent column density of $H_2$ per cm$^{-2}$.

For the low resolution data we beam match the THINGS 21-cm maps to the $13.3''$ resolution of the HERACLES maps. We do not convolve the high-resolution CO maps at all, but work with them at their native resolutions. Also note that we do not apply any inclination corrections to these measurements. We compare to luminosities measured in apertures, also without inclination correction. We consider the total mean column density through the aperture to represent the relevant quantity for attenuation.

\section{Analysis}
\label{sec:analysis}

We focus our analysis on individual apertures. For each aperture, we subtract a local background from the \pabeta\ and \halpha\ maps and infer the luminosity of each line in the aperture. Using these measurements, we infer the attenuation affecting \halpha\ in each aperture and calculate the attenuation-corrected ionizing photon production rate. We chose the aperture diameters of 2\arcsec{} and 3\arcsec{} based on the resolution of the seeing-limited \halpha{} data ($\lesssim 2''$) and the 33 GHz VLA images ($\sim 3''$).

Our original intention was to also perform a pixel-by-pixel analysis that considers all emission. Unfortunately, the final mosaicked \pabeta\ images show too much low level background variation to allow large scale integration or convolution to low resolution as in \citet{LI13}. In working with the individual apertures, we subtract a local background determined from the same region for both \pabeta\ and \halpha .

The zero point for the \halpha\ appears more stable, and we can achieve robust measurements at coarser resolution. For each aperture, we also measure the IR-to-H$\alpha$ color and gas column densities at a series of coarser resolutions. When we do this, we first convolve the IR, \halpha , or neutral gas map. Then we measure the mean value of this low resolution map within each $2''$ or $3''$ apertures. In the limit where the new beam is large compared to the $2''$ aperture, this approaches sampling the low resolution map at the center of each aperture.

\subsection{Aperture placement}
\label{sec:apertures}

We place apertures in each location where we would expect to detect both \pabeta{} and \halpha{} in the absence of dust. To set this criteria, we first measure the rms noise in the \pabeta\ map,  $\sigma_{PB}$ . Then, we scale $\sigma_{PB}$ (1.46\e{-6} \intunits in NGC~5194 and 5.63\e{-7} \intunits in NGC~6946) to a corresponding \halpha\ intensity assuming no dust and case B recombination, i.e., a ratio of 16.7 in NGC~5194 and 17.5 in NGC~6946.

\begin{figure*}
    \centering
    \includegraphics[width = 0.45\textwidth]{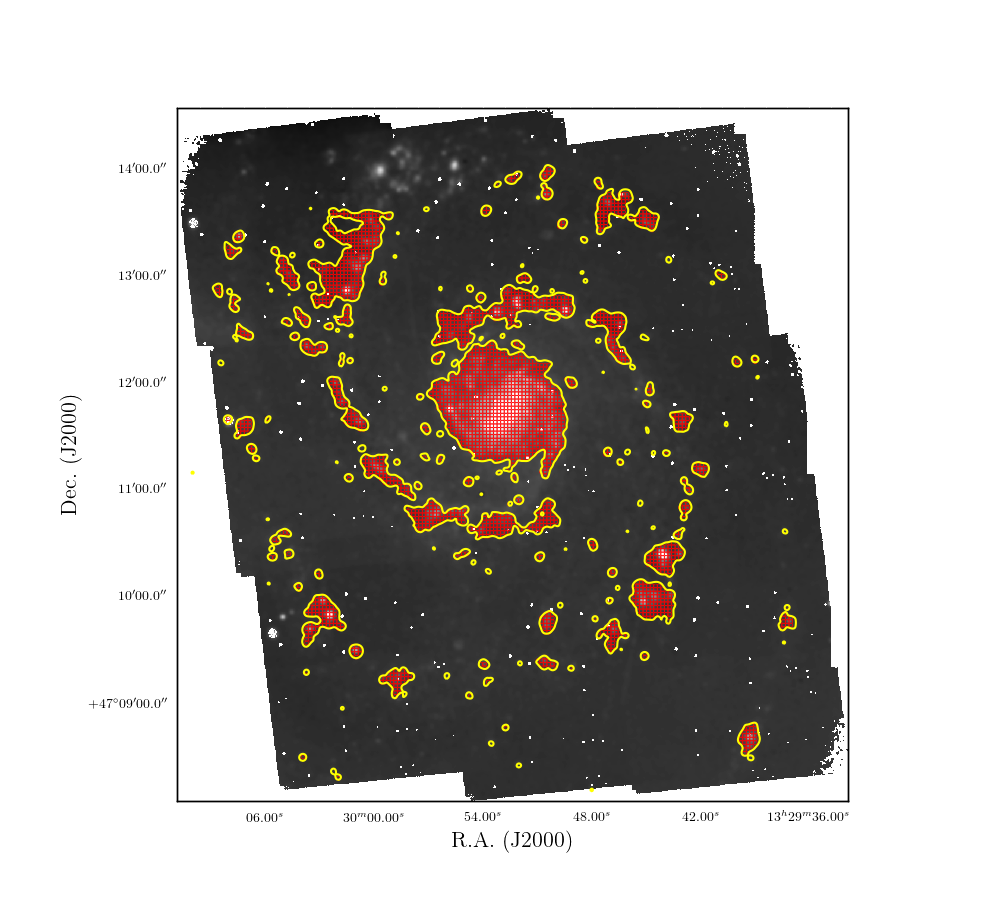}
    \includegraphics[width = 0.45\textwidth]{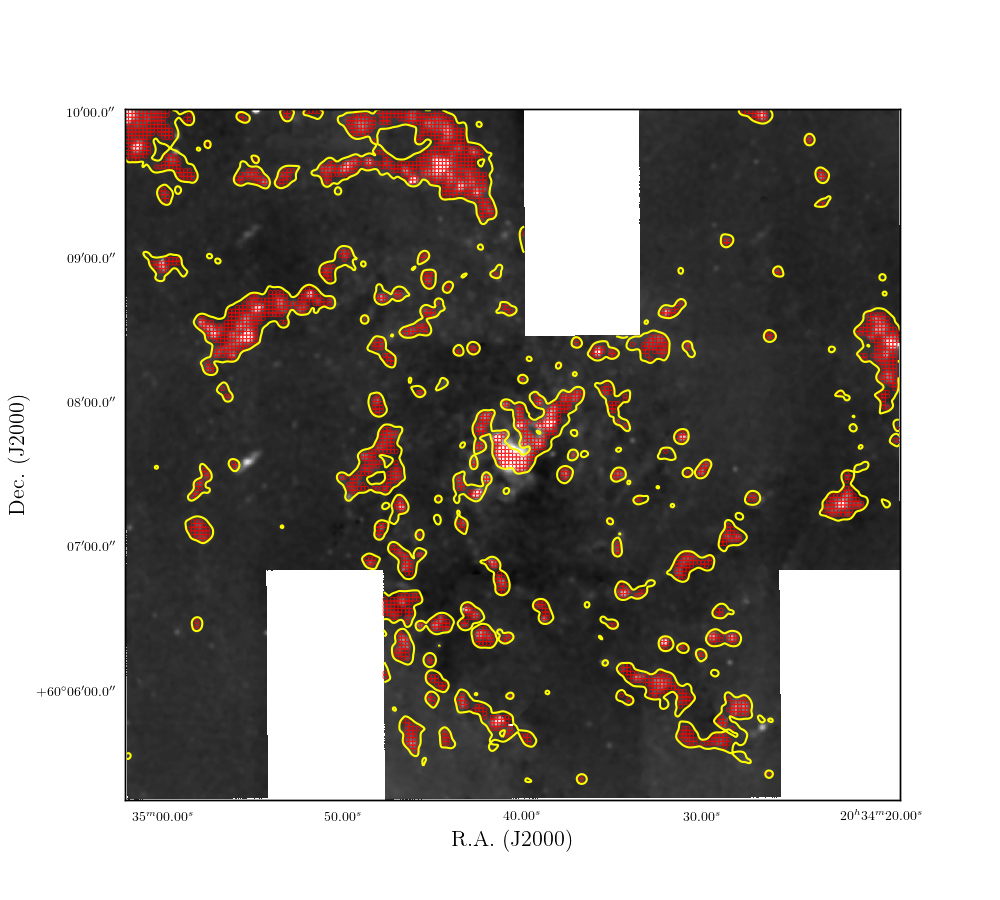}
    \caption{\textbf{\pabeta{} images overlayed with \halpha{} contours in yellow which we use to place apertures and the apertures in red. } The contours are placed where \halpha{} is equal to the intrinsic ratio of  \halpha{} to \pabeta (16.7 in NGC~5194 and 17.5 in NGC~6946) multiplied by 1 $\sigma_{PB}$. The 2\arcsec{} apertures are plotted in red. This aperture placement focuses our analysis on regions of the galaxy with bright \halpha{}. Our 1$\sigma$ threshold in \halpha{} causes us to miss a handful of heavily attenuated regions in the center of NGC~6946, but these will not substantially affect our analysis. We do not find any similar set of missed regions in NGC~5194.}
    \label{fig:Contours galaxies}
\end{figure*}

We place apertures in regions that exceed this threshold intensity in the \halpha\ map. That is, we place apertures everywhere that we would expect to detect \pabeta\ at S/N$>1$ based on the \halpha\ map if there were no dust. Dust is certainly present, and any attenuation makes the \pabeta\ brighter compared to \halpha. This makes this cut fairly conservative. To verify this, Figure \ref{fig:Contours galaxies} shows our threshold intensity contour, in yellow, on the \pabeta\ images. This approach selects most of the bright regions in both galaxies.

In regions above the \halpha\ threshold, we place 2\arcsec{} diameter apertures. We use a square grid with adjacent apertures overlapping in area by 10\%. This 2\arcsec{} scale corresponds to a physical size of 73 pc at the distance of NGC~6946 and 83 pc at the distance of NGC~5194. The aperture placement is shown in Figure \ref{fig:Contours galaxies}. 

For NGC~5194 we repeat this exercise using 3\arcsec{} diameter apertures for comparison to the 33~GHz radio continuum data of \citet{QUEREJETA2019} and the H$\alpha$ spectroscopy of \citet{BLANC09}.  NGC~6946 lacks these comparison data, and we do not make 3\arcsec\ apertures for this galaxy. 

By using an \halpha{} threshold we do miss some very extinguished regions in the center of NGC~6946. These regions are absent from the \halpha{} image, but bright in our \pabeta{} map. If we select on \pabeta{} instead of \halpha{} (we make a one sigma cut in \pabeta), we increase the number of apertures by $\sim 2.5\%$. As expected, the additional apertures appear in the inner, heavily extinguished part of the galaxy. In addition to these apertures appearing bright in \pabeta{}, they also appear bright in the archival infrared continuum maps that we utilize (see Section \ref{sec:data}).

That the apertures appear bright in both \pabeta{} and longer wavelengths dust continuum leads us to believe that these regions are real. However, for this paper we focus only on apertures that meet our \halpha{} threshold. In NGC~5194, we do not find any analogous cases where we would select an aperture based on \pabeta{} but it does not appear visible in \halpha{} emission. In this paper, we focus on studying \Ahalpha, but we note that studying these heavily embedded sources in the inner part of NGC~6946 represents a good future application for our data.

In NGC~5194 we place a total of 2,075 apertures in the 2\arcsec{} map and 588 apertures in the 3\arcsec{} map. In NGC~6946 we place a total of 1,822 2\arcsec\ diameter apertures.

\subsection{Measurements in apertures}\label{sec:measurements in apertures}

We subtract a local background from each aperture. We place a circular annulus around each aperture that has an inner radius of 4\arcsec{} and an outer radius of 20\arcsec. We use the median value of each annuli as our local background to subtract from the corresponding aperture. For most cases these annuli are dominated by nearby empty sky instead of nearby bright emission. However, in the center of NGC~5194 these annuli are dominated by bright emission which caused us to over-subtract the background. In the center 3kpc (the deprojected galactocentric radius is calculated assuming the distance and orientation in Table \ref{tab:Adopted Properties}) of NGC~5194, we used large annuli to ensure that we accurately calculated the local background. These large annuli have an inner radius of 14\arcsec{} and an outer radius of 125\arcsec. 

After subtracting the background, we integrate emission from each aperture to calculate the total line flux. Scaling the line flux by the distances in Table \ref{tab:Adopted Properties}, we calculate $L(\pabeta)$ and $L(\halpha)$.

The 2\arcsec{} diameter of the apertures matches the FWHM of the (convolved, Gaussian) PSF of both images. To account for emission associated with the part of the PSF outside the aperture, we apply an aperture correction to all measured fluxes. For a Gaussian PSF and a point source, an aperture with diameter equal the FWHM of the PSF requires an aperture correction factor of 2.0 to recover the correct flux. The aperture correction for extended sources may differ from this value. As long as \halpha{} and \pabeta{} show the same structure, any inaccuracy in the aperture correction will affect both lines in the same way. Therefore, we do not expect our choice of aperture correction to affect our derived extinction estimates.

These measurements, along with the inferred \Ahalpha\ and our best estimate uncertainties, are tabulated in Tables \ref{tab: aperture_table_6946} and \ref{tab: aperture_table_5194}.

For each aperture, we also note: (1) The galactocentric radius, calculated assuming the distance, orientation, and inclination in Table \ref{tab:Adopted Properties}. (2) The IR-to-\halpha\ colors for IR maps at 8, 12, 24, 70, and (only for NGC~6946) 100$\mu$m. Following Section \ref{sec:Halpha+IR}, these ratios are derived at lower resolution, with the exact resolution set by the IR map in question. (3) Atomic ({\sc Hi}), molecular (H$_2$), and total ({\sc Hi}+H$_2$) gas column densities at several resolutions, as described in Section \ref{sec:column}. For the gas and IR-to-\halpha\ colors, we record the mean column density or color in the aperture instead of the aperture sum. This is due to concerns about convolving our \pabeta{} maps to match the large IR PSFs and is explained further in Section \ref{sec:Halpha+IR}.

\begin{deluxetable*}{lcccccccc}
\tabletypesize{\scriptsize}
\tablecaption{Measured Line Luminosities and Estimated Attenuation in NGC~6946 \label{tab: aperture_table_6946}}
\tablewidth{\linewidth}
\tablehead{
\colhead{ID} &
\colhead{R.A.} &
\colhead{Dec.} &
\colhead{L(H$\alpha$)} &
\colhead{$\sigma$(L(H$\alpha$))} &
\colhead{L(Pa$\beta$)} &
\colhead{$\sigma$(L(Pa$\beta$))} &
\colhead{A(H$\alpha$)} &
\colhead{$\sigma$(A(H$\alpha$))} \\
\colhead{} &
\colhead{(deg.)} &
\colhead{(deg.)} &
\colhead{(\lunits )} &
\colhead{(\lunits )} &
\colhead{(\lunits )} &
\colhead{(\lunits )} &
\colhead{(mag)} &
\colhead{(mag)}
}
\startdata
1 & 308.8488 & 60.1622 & 2.1007e+37 & 1.5664e+36 & 1.8951e+36 & 1.0943e+36 & 0.7256 & 0.0852 \\
2 & 308.8493 & 60.1626 & 2.7726e+37 & 1.4515e+36 & 2.8458e+36 & 1.7624e+35 & 0.9294 & 0.0583 \\
3 & 308.8498 & 60.163 & 3.1386e+37 & 1.6424e+36 & 7.0358e+36 & 3.8189e+35 & 2.1635 & 0.0325 \\
4 & 308.8504 & 60.1634 & 1.6762e+37 & 8.8059e+35 & 1.6436e+36 & 1.2576e+35 & 0.8574 & 0.0978 \\
5 & 308.8525 & 60.1652 & 1.1600e+37 & 6.7633e+35 & 7.3008e+35 & 6.3087e+35 & 0.1569 & 0.7366 \\
6 & 308.8535 & 60.166 & 1.6622e+38 & 8.7813e+36 & 9.9786e+36 & 1.6111e+36 & 0.0822 & 0.3975 \\
7 & 308.854 & 60.1664 & 3.0860e+38 & 1.6229e+37 & 2.2777e+37 & 2.8773e+36 & 0.4084 & 0.2880 \\
8 & 308.8546 & 60.1669 & 3.3841e+38 & 1.7715e+37 & 2.5994e+37 & 3.2914e+36 & 0.4716 & 0.2926 \\
9 & 308.8474 & 60.162 & 1.3413e+37 & 1.2144e+36 & 1.2973e+36 & 3.4305e+35 & 0.8357 & 1.4941 \\
10 & 308.8479 & 60.1624 & 3.4726e+37 & 2.0743e+36 & 3.7959e+36 & 5.5747e+35 & 1.029 & 0.3952 \\
\nodata 
\enddata
\tablecomments{Luminosities and estimated attenuation for 2\arcsec\ diameter apertures placed in NGC~6946. Luminosities assume the distances in Table \ref{tab:Adopted Properties}. The $\sigma$ quantities are our 1$\sigma$  uncertainties. Details on how these uncertainties were calculated are described in Sec. \ref{sec:uncertainties} This table is a stub. We show a preview of the first 10 lines here. The full machine-readable table is available for download online.}
\end{deluxetable*}

\begin{deluxetable*}{lcccccccc}
\tabletypesize{\scriptsize}
\tablecaption{Measured Line Luminosities and Estimated Attenuation in NGC~5194 \label{tab: aperture_table_5194}}
\tablewidth{0pt}
\tablehead{
\colhead{ID} &
\colhead{R.A.} &
\colhead{Dec.} &
\colhead{L(H$\alpha$)} &
\colhead{$\sigma$(L(H$\alpha$))} &
\colhead{L(Pa$\beta$)} &
\colhead{$\sigma$(L(Pa$\beta$))} &
\colhead{A(H$\alpha$)} &
\colhead{$\sigma$(A(H$\alpha$))} \\
\colhead{} &
\colhead{(deg.)} &
\colhead{(deg.)} &
\colhead{(\lunits )} &
\colhead{(\lunits )} &
\colhead{(\lunits )} &
\colhead{(\lunits )} &
\colhead{(mag)} &
\colhead{(mag)}
}
\startdata
1 & 202.5333 & 47.22 & 2.0723e+37 & 1.0825e+36 & 1.3358e+36 & 1.1488e+35 & 0.1209 & 0.1381 \\
2 & 202.5333 & 47.2205 & 1.9726e+37 & 1.0317e+36 & 1.9419e+36 & 2.1371e+35 & 0.7899 & 0.1857 \\
3 & 202.5333 & 47.2211 & 1.6210e+37 & 8.5108e+35 & 1.5476e+36 & 2.7520e+35 & 0.7414 & 0.3206 \\
4 & 202.5325 & 47.2116 & 1.5544e+37 & 8.1388e+35 & 1.7330e+36 & 1.4290e+35 & 0.9865 & 0.1034 \\
5 & 202.5325 & 47.2122 & 1.2588e+37 & 6.6195e+35 & 1.4802e+36 & 1.2302e+35 & 1.0707 & 0.1215 \\
6 & 202.5325 & 47.2127 & 1.4428e+37 & 7.5649e+35 & 1.4014e+36 & 1.3301e+35 & 0.7687 & 0.1345 \\
7 & 202.5325 & 47.22 & 2.5569e+37 & 1.3325e+36 & 2.2734e+36 & 1.7022e+35 & 0.629 & 0.0870 \\
8 & 202.5325 & 47.2205 & 2.3530e+37 & 1.2274e+36 & 3.7515e+36 & 3.4611e+35 & 1.5516 & 0.1124 \\
9 & 202.5325 & 47.2211 & 1.2190e+37 & 6.4402e+35 & 2.4065e+36 & 2.2078e+35 & 1.8891 & 0.1122 \\
10 & 202.5317 & 47.2116 & 1.8666e+37 & 9.7473e+35 & 1.7931e+36 & 2.1323e+35 & 0.7512 & 0.1550 \\
\nodata
\enddata

\tablecomments{Luminosities and estimated attenuation for 2\arcsec\ diameter apertures placed in NGC~5194. Luminosities assume the distances in Table \ref{tab:Adopted Properties}. The $\sigma$ quantities are our 1$\sigma$  uncertainties. Details on how these uncertainties were calculated are described in Sec. \ref{sec:uncertainties}. This table is a stub. We show a preview of the first 10 lines here. The full machine-readable table is available for download online.}
\end{deluxetable*}

\subsection{Attenuation estimates}
\label{sec: Extinction Measurements}

Assuming Case B recombination, the ratio of \halpha{} emission to \pabeta{} emission in any given area should be 17.5 (at T = 10,000K and $\rho$ = 1000 cm$^{-3}$) in the absence of dust \citep{HUMMER87}. This number is relatively insensitive to the adopted density and temperature conditions. If we allow the density to range from 300 to 3,000 cm$^{-3}$ and the temperature to range from 5,000 to 15,000 K the expected ratios range from 16.5 to 18.3.

Dust attenuates \halpha{} emission more than \pabeta{} emission because \halpha{} is emitted at a shorter wavelength. In practice, this leads to a lower than expected \halpha -to-\pabeta\ ratio. Each magnitude of selective attenuation, e.g., expressed as $E(B-V)$, implies more \halpha\ attenuation than \pabeta\ attenuation, driving the ratio progressively lower. As a result, the difference between the expected and observed line ratio, along with an adopted extinction curve, implies some total attenuation along the line of sight.

This calculation relies on the ratios among attenuation at different wavelengths, here \halpha\ and \pabeta , remaining fixed. This is a standard assumption in the field, and here we assume that such an ``attenuation law'' holds across our target galaxies. We adopt the attenuation curve from \citet{CCM89} and mostly express the total attenuation in terms of \halpha{}. From \citet{CCM89} we note the conversions $
1\ mag\ A_{H\alpha} = 0.31\ mag\ A_{Pa\beta} = 0.37\ mag\ E(B-V).$

Assuming an extinction curve, \Ahalpha{} is related to the ratio of \halpha{} to \pabeta{} via

\begin{equation}\label{eq:fluxrelation}
    \log_{10} \left(\frac{L_{H\alpha}}{L_{Pa\beta}}\right)_{obs} = \log_{10}(R_{\rm int}) \cdot \frac{A(H\alpha)}{2.5}\left(\frac{k(Pa\beta)}{k(H\alpha)} - 1\right)
\end{equation}

\noindent Here $k(H\alpha)$ and $k(Pa\beta)$ correspond to the adopted reddening curves evaluated at the wavelengths of \halpha{} and \pabeta{} . For our fiducial adopted \citet{CCM89} extinction law, $k(H\alpha) = 2.68$ and $k(Pa\beta)=0.84$. $R_{\rm int}$ refers to the intrinsic ratio of \halpha\ to \pabeta\ in the absence of attenuation. For NGC~5194, we adopt $R_{\rm int}=16.7$, reflecting the temperature of T = 6300($\pm$500)K found by \citet{CROXALL15}, which reduces the expected ratio of \halpha{} to \pabeta{} from 17.5 (for $T=10,000$~K) to 16.7 \citep{PYNEB}. For NGC~6946 we assume a temperature of $T=10,000$~K and adopt $R_{\rm int} = 17.5$.

Rearranging Eq. \ref{eq:fluxrelation}, \Ahalpha takes the form of

\begin{equation}\label{eq:Aha}
    A_{H\alpha} = \frac{k(H\alpha)}{k(Pa\beta) -k(H\alpha)} \cdot 2.5 \log\frac{(H\alpha / Pa\beta)_{obs}}{(H\alpha / Pa\beta)_{int}} .
\end{equation}

In addition to a single, fixed extinction law, Equation \ref{eq:Aha} assumes a simple geometry in which all light is processed through a foreground screen of dust. In a realistic star-forming galaxy, the geometry may be more complex. One commonly-considered more complex scenario is a ``mixture model.'' In this scenario, dust is evenly mixed with {\sc Hii} regions. Some regions will experience lower total attenuation, while others experience higher total attenuation. The observed line ratio reflects the sum of light from all regions. We return to the effects of geometry in Section \ref{sec:Gas v Ahalpha}.

\begin{figure*}
    \centering
    \includegraphics[width = 0.45\linewidth]{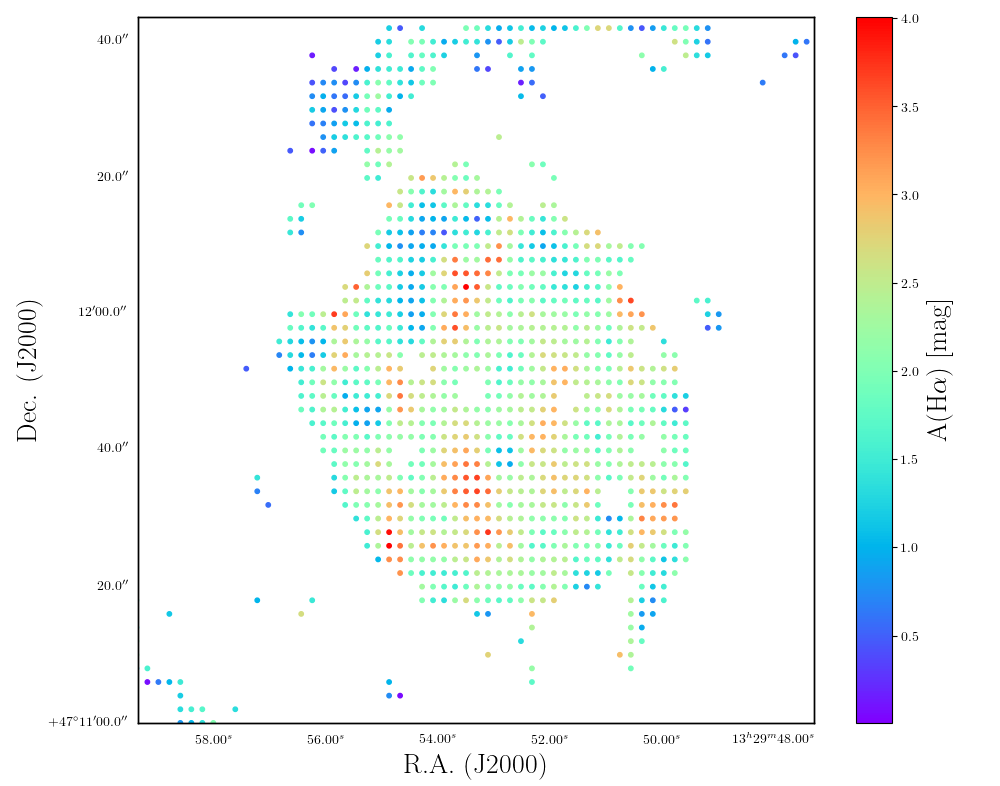}
    \includegraphics[width = 0.45\linewidth]{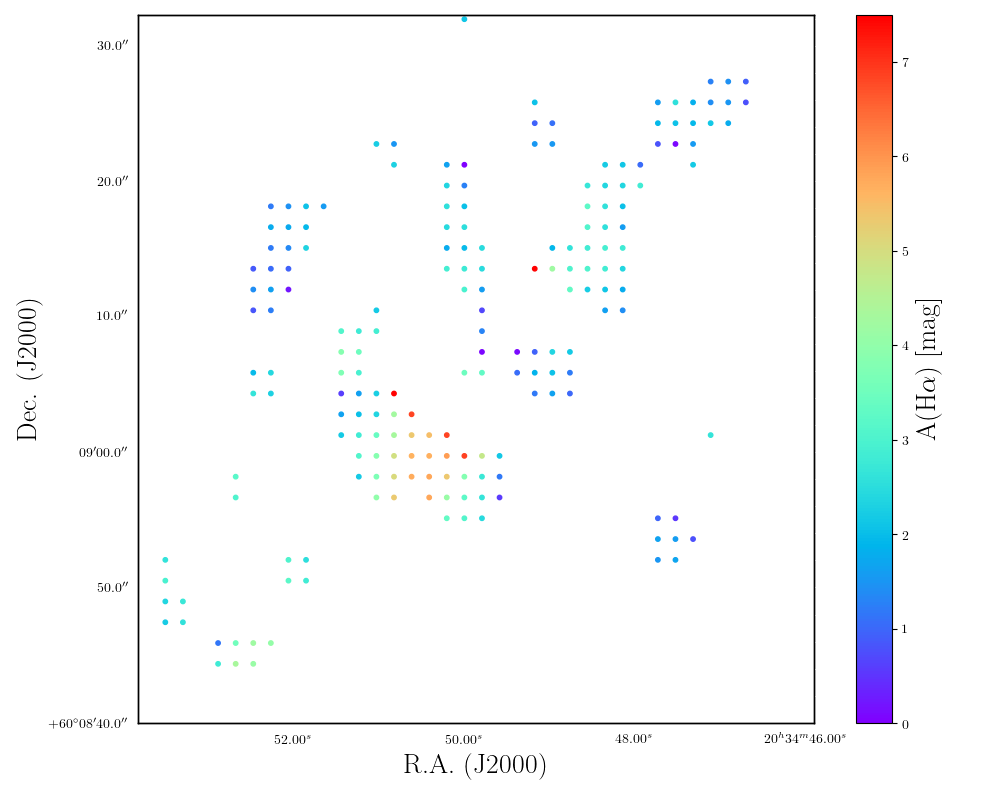}\\
    \includegraphics[width = 0.45\linewidth]{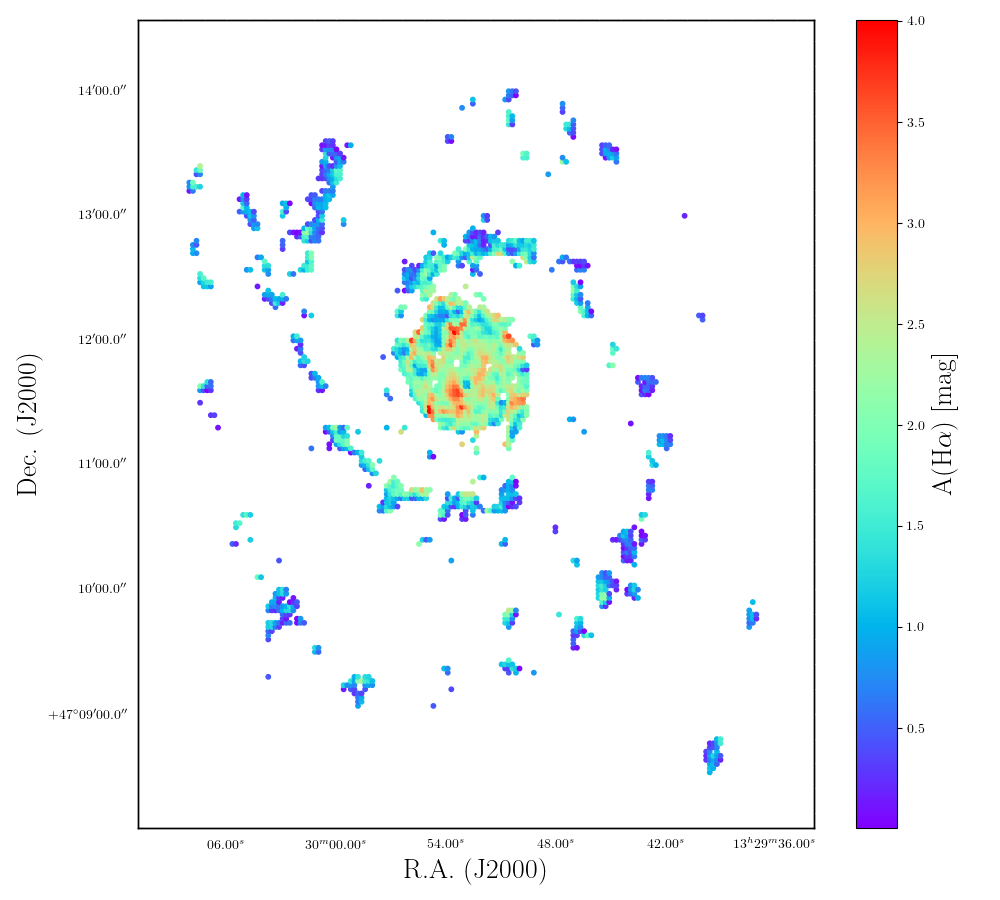}
    \includegraphics[width = 0.45\linewidth]{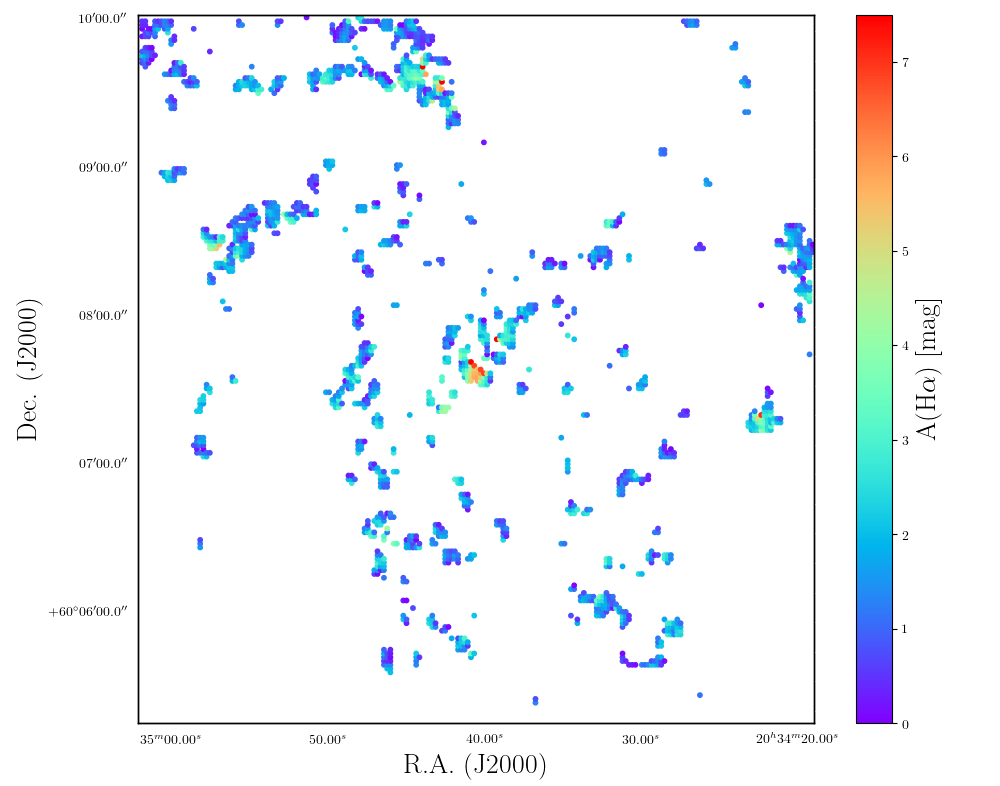}
    \caption{\textbf{Spatial distribution of \Ahalpha{} in NGC~5194 and NGC~6946.} We calculate the value of \Ahalpha{} in each aperture and plot the value of each aperture for NGC~5194 (left) and NGC~6946 (right). The bottom two panels cover the full galaxy. The top two panels are zoomed in to show the centers of the two galaxies. }
    \label{fig:Aha_spatial}
\end{figure*}

\subsection{Ionizing photon production rate}
\label{sec:q0}

\halpha{} traces the recombination of hydrogen atoms. Photoionizations balance recombinations in {\sc Hii} regions. Therefore the \halpha{} luminosity, once corrected for attenuation, directly traces the production rate of ionizing photons.

In this paper we express the rate of ionizing photon production in terms of the attenuation-corrected \halpha\ luminosity, $L_{corr}(H\alpha)$. Because we consider individual apertures on small scales, we do not cast our results in terms of the star formation rate. But for reference and comparison to other work, we note the relationship between SFR and $L_{corr}(H\alpha)$ given in  \citet{KENNICUTT12}:

\begin{equation}
\label{eq: sfrhaeq}
SFR(M_{\odot} \ yr^{-1}) = 5.4 \times 10^{-42} L_{corr}(H\alpha)(erg \ s^{-1}).
\end{equation}

\noindent This relation was derived by \citet{MURPHY11} using Starburst99 \citep{LEITHERER99} model calculations for continuous star formation and a Kroupa initial mass function \citep{KROUPA01} over a stellar mass range of 0.1-100M$_{\odot}$.

This paper focuses on the effect of attenuation on recombination line emission from {\sc Hii} regions, but a full accounting of ionizing photons would also require addressing direct absorption of ionizing photons by dust or leakage of ionizing photons. Both effects occur. For example see \citet{BINDER18} for a study showing dust absorption of ionizing photons. Meanwhile there is extensive evidence showing diffuse ionized gas in our Galaxy and others. Photons leaked from {\sc Hii} regions represent the likely ionizing photon source for much of this diffuse ionized gas \citep[e.g.,][]{HAFFNER09}.

We compare our results for NGC~5194 to the 33~GHz free-free radio continuum map of \citet{QUEREJETA2019}. This emission traces random close encounters between protons and electrons, so that its dependence is similar to \halpha , but it is largely unaffected by dust attenuation.

We utilize Equations 10 and 11 of \citet{MURPHY11} to convert between 33~GHz free-free luminosity, $L_\nu^{T}$, and $L_{corr}(H\alpha)$. These equations relate the ionizing photon production rate, $Q(H^0)$, to the free-free luminosity via

\begin{equation}
    \begin{split}
    \left[\frac{Q(H^0)}{s^{-1}}\right] = 6.3\times 10^{25}\left(\frac{T_e}{10^{4} K}\right)^{-0.45} \left(\frac{\nu}{GHz}\right)^{0.1} \\
    \times \left(\frac{L_{\nu}^{T}}{erg\ s^{-1}\ Hz^{-1}}\right),
    \end{split}
\end{equation}

\noindent where for us $\nu = 33$~GHz and $T_e = 6,300$~K for NGC~5194 following \citet{CROXALL15}. Then, as given in \citet{MURPHY11}, $Q(H^0)$ relates to the star formation rate via,

\begin{equation}\label{eq: sfrfreefreeeq}
    \begin{split}
    \left(\frac{SFR^{T}_{\nu}} {M_{\odot} yr^{-1}}\right) = 4.6 \times 10^{-28} \ \left(\frac{T_e}{10^{4} K}\right)^{-0.45} \left(\frac{\nu}{GHz}\right)^{0.1} \\
    \times \left(\frac{L_{\nu}^{T}}{erg\ s^{-1}\ Hz^{-1}}\right).
    \end{split}
\end{equation}

After we calculate $SFR^{T}_{\nu}$, we use Equation \ref{eq: sfrhaeq} to convert to the equivalent $L_{corr}(H\alpha)$. We emphasize that the SFR here only represents a convenient variable to convert from $Q(H^0)$ to $L_{corr}(H\alpha)$. That is, we only use these equations together to convert 33~GHz to the predicted corresponding $L_{corr}(H\alpha)$.

\subsection{Uncertainties}
\label{sec:uncertainties}

Our \pabeta{} and \halpha\ measurements are affected by both statistical noise and uncertainty in the background level. Our estimates of \Ahalpha\ also depend on the assumed extinction curve, as well as the density and temperature, which set the intrinsic \halpha-to-\pabeta\ ratio.  

\textbf{Statistical noise:} We estimate the statistical noise in our data by placing apertures in apparently emission-free regions of each map. We placed $\sim 10,000$ apertures in each galaxy. In the appendix, we show the distribution of luminosities measured in these blank-sky apertures. As expected, the distribution appears centered on or near $\sim 0$~erg~s$^{-1}$, confirming that our background subtraction appears to be working well. NGC~5194's histogram is centered slightly less than zero at -7.52$\times 10^{-34}~erg~s^{-1}$ but this is controlled for in our analysis by local background subtraction. We adopt the rms scatter in the luminosity measured for these apparently empty apertures as our estimate of the statistical noise.

For NGC~5194 we estimate statistical noise of $\sigma_{Pa\beta} = 6.3 \times 10^{35}$~\lunits\ for \pabeta\ and $\sigma_{H\alpha} = 6.8 \times 10^{35}$~\lunits\ for \halpha . For NGC~6946 we estimate statistical noise of $\sigma_{Pa\beta} = 8.7 \times 10^{34}$~\lunits\ for \pabeta\ and $\sigma_{H\alpha} = 9.1 \times 10^{34}$~\lunits\ for \halpha .

\textbf{Zero point uncertainty:} We used an empirical approach to continuum subtraction, bootstrapping the appropriate ratio to translate OFF emission to a continuum estimate from the data themselves. This process has some associated uncertainty. We estimate the magnitude of this uncertainty by making six maps that range from being visibly slightly over-subtracted to slightly under-subtracted. 

We allow our Monte Carlo error estimation (described below) to randomly choose between one of these maps. This allows us to quantify the error associated with our continuum subtraction. Based on this exercise, we estimate the rms uncertainty in the overall background level to be 7.00\e{36}~\lunits\ in \pabeta\ and 3.00\e{37}~\lunits\ in \halpha{} for NGC~5194. In NGC~6946 we estimate the degree of rms uncertainty in the overall background level to be 1.11\e{36}~\lunits\ in \pabeta\ and 2.12\e{37}~\lunits\ in \halpha.

Note that unlike the statistical noise, this uncertain background will affect all (or at least many) apertures in tandem.

\textbf{Combined uncertainty estimates:} We estimate the total uncertainty on each measured luminosity and calculated \Ahalpha{} value by running a Monte Carlo simulation. 

We treat our measurements as the ``true'' values. Then we add statistical noise, apply other sources of uncertainty, and repeat our calculations. We aim to allow any variables that might affect the final value of \Ahalpha{} to vary across a realistic range. 

Specifically, we do the following, beginning with our true measurements:

\begin{enumerate}
\item Add normally distributed statistical noise to each L(\halpha) and L(\pabeta) value with the magnitude noted above.
\item Perturb the zero point of all L(\halpha) and L(\pabeta) by adding or subtracting the photometric calibration error. This error can vary between $5-7\%$ and we therefore we choose a random error of either $5\%$, $6\%$, and $7\%$.
\item Choose randomly from six \pabeta{} maps where the continuum is visibly over and under subtracted to various degrees.
\item Allow the temperature and density used to calculate the intrinsic line ratio to vary.  For the temperature, we choose a random temperature from a Gaussian distribution with FWHM 1,000~K centered on the median temperature of the galaxy [6,300~K for NGC~5194 \citet{QUEREJETA2019} and 10,000~K for NGC~6946 \citet{MURPHY11}]. For the density, we any allow density between 100-10,000 $cm^{-3}$ in linear space.
\item Randomly choose between the \cite{CCM89} extinction curve and the \citet{FITZPATRICK99} extinction curve. 
\end{enumerate}

We ran 1,000 simulations. We adopt the rms scatter in the measured luminosities and \Ahalpha{} as our best-estimate uncertainty. For \Ahalpha , we find rms scatter of 0.35 mag in NGC~5194 and 0.20 mag in NGC~6946. In Tables \ref{tab: aperture_table_5194} and \ref{tab: aperture_table_6946} we use this calculation to report 1$\sigma$ values for each reported quantity.

\subsection{Comparison to literature measurements}
\label{sec: compare lit}

There has not been previous wide-field \pabeta\ imaging of either of our targets, but both NGC~5194 and NGC~6946 have multiple previous estimates of \Ahalpha . NGC~5194, in particular, has been a key target for calibrating recipes to correct \halpha\ for the effects of attenuation.

\subsubsection{NGC~5194}

\begin{figure*}
\centering
\includegraphics[width = .45\textwidth]{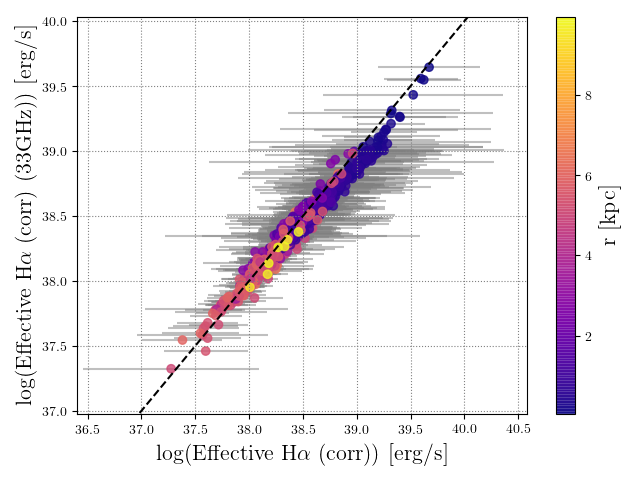}
\includegraphics[width =.45\linewidth]{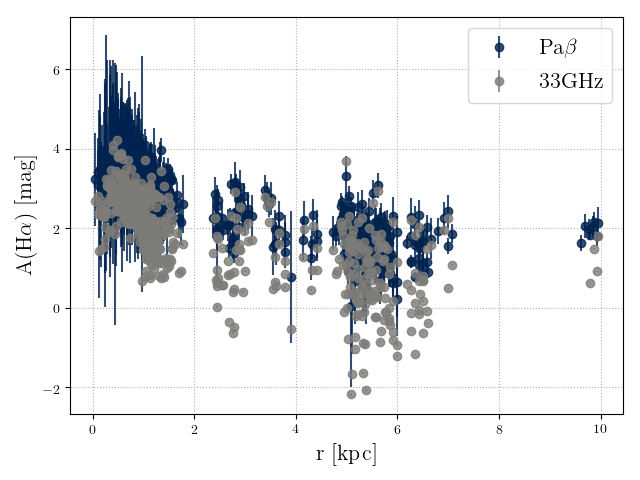}
\caption{\textbf{Comparison of our measurements to 33~GHz radio continuum.} \textbf{Left:} Extinction-corrected \halpha{} luminosity estimated from 33~GHz free-free radio continuum emission \citep{QUEREJETA2019} as a function of our attenuation-corrected \halpha{} emission estimates from combining \pabeta\ and \halpha . When estimating the extinction corrected \halpha{} luminosity, we assume that 70\% of the 33 GHz emission comes from free-free emission. Each point corresponds to measurements for a 3\arcsec{} aperture in NGC~5194, with apertures color-coded by radius. The dashed line shows the one-to-one line expected for perfect agreement. Across all apertures, the median ratio of 33~GHz-based \halpha{} luminosity to our estimate corrected \halpha{} luminosity is 0.88 with 0.07 dex rms scatter. This shows good overall agreement between these t. \textbf{Right}: Two estimates of H$\alpha$ attenuation, \Ahalpha , as a function of galactocentric radius. In black, we show \Ahalpha\ estimated by comparing \pabeta{} and \halpha{}. In gray, we show \Ahalpha\ estimated by comparing \halpha{} to 33~GHz emission. As shown in the panel on the left, the \pabeta{}-to-\halpha{} approach shows overall good agreement but yields slightly higher attenuation than the \halpha{} to radio continuum. The biggest differences come in the inner part of the galaxy.}
\label{fig:ionizng photons, freefreevPaB, 5194}
\end{figure*}

\textbf{Large area 33~GHz map:} The recent 33~GHz radio continuum map by \citet{QUEREJETA2019} covers a large part of NGC~5194. We measure 33~GHz luminosities in matched apertures from the 33~GHz map and our own data, following the same local background subtraction method detailed in \ref{sec:apertures}. In this paper we assume that the free-free makes up  70\% of the total 33~GHz radio emission across all apertures. This might overestimate the free-free emission in the central 3 kpc which might be affected by NGC~5194's central AGN (see Section \ref{sec:data}). Then, we re-express the free-free luminosity as attenuation-corrected \halpha\ following Section \ref{sec:q0}.

In Figure \ref{fig:ionizng photons, freefreevPaB, 5194} we compare results from the \citet{QUEREJETA2019} free-free map to the corrected \halpha{} from our \pabeta{} based measurements. We compare the extinction-corrected \halpha{} luminosities from the two methods in the left panel and \Ahalpha{} in the right panel. We obtain an effective attenuation for the 33~GHz data by comparing it to our \halpha{} image and assuming the 33~GHz data is not attenuated by dust.

Overall the two approaches to estimate \Ahalpha{} agree fairly well. Considering the corrected \halpha{} luminosity, the value estimated from free-free emission is median 0.88 times that estimated from \pabeta{} and \halpha{}, with rms scatter in the ratio of 0.07 dex. If the free-free fraction were allowed to shift to $\sim 0.8$, the two data sets would show a median ratio of $\sim 1$.

Figure \ref{fig:ionizng photons, freefreevPaB, 5194} also shows \Ahalpha{} as a function of galactocentric radius, plotting results for both the free-free and \pabeta{}-based approaches. The figure shows that in the central 2~kpc the \Ahalpha{} from our \pabeta{} approach is higher than that from the free-free approach. At larger radii, the free-free emission predicts slightly larger \Ahalpha{} than the \pabeta{}. Again in terms of corrected \halpha{ }luminosity, the median ratio (free-free-based estimate to \pabeta-based-estimate) inside 2~kpc is 0.63 with a rms scatter in the ratio 0.99 mag. The median ratio in the outer 3-9~kpc is 1.32 with a rms scatter in  the ratio of 0.19 mag.

We expect the two approaches to yield the same result, so this comparison suggests a $\sim \pm 30\%$ uncertainty on the final attenuation-corrected luminosities. Some of these differences might result from flux calibration uncertainties and uncertainties in background subtraction. The fraction of the radio emission arising from free-free emission also represents an uncertainty. We adopted a constant free-free fraction (see Section \ref{sec:data}), but this quantity likely might vary spatially \citep[e.g., see discussion in][]{QUEREJETA2019}, which might create some of the gradient.

In the very center of the galaxy, NGC~5194's AGN likely plays a role. The likely impact of the AGN will be to contribute additional synchrotron emission. In this case the fraction of the radio coming from free-free emission will be smaller at the center of the galaxy than in the surrounding regions. There is some evidence for this already in Figure \ref{fig:ionizng photons, freefreevPaB, 5194}. In the left panel, the highest luminosity points correspond to the center. These curve up towards equality, showing higher 33~GHz emission relative to the surrounding (blue and purple) points. \citet{QUEREJETA2019} estimated the thermal fraction from the spectral index and found that it is smaller in the center. In fact the area where the thermal fraction is smaller precisely coincides with the area of the AGN radio plasma jet as seen at other radio wavelengths. This is nicely confirmed in Fig. \ref{fig:ionizng photons, freefreevPaB, 5194}.

\textbf{Recombination line measurements from the inner 2~kpc:} The inner part of NGC~5194 has been observed in recombination line emission and radio free-free emission many times, with attenuation estimated in studies by \citet{CALZETTI05}, \citet{KENNICUTT07}, \citet{CALZETTI07}, \citet{BLANC09}, and \citet{QUEREJETA2019} among others. 

\begin{figure}
     \centering
     \includegraphics[width = .45\textwidth]{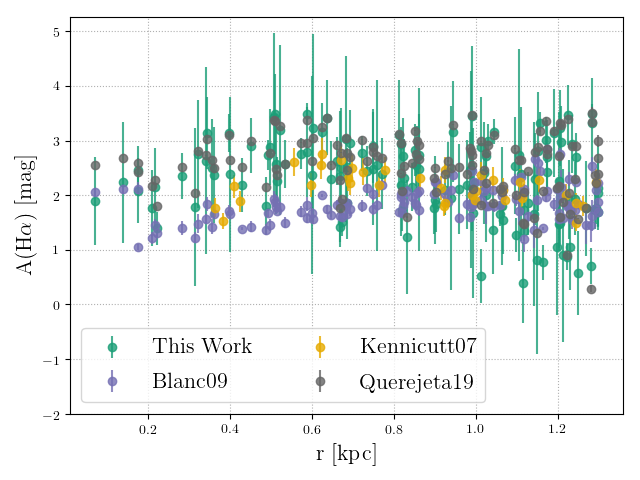}
     \caption{\textbf{Comparisons to past literature in NGC~5194.} Plotted are \Ahalpha{} measurements at aperture locations and sizes from \citet{BLANC09}. Our measurements agree well with both \citet{KENNICUTT07} and attenuation measurements derived using the 33~GHz map from \citet{QUEREJETA2019} and \halpha{} assuming the 33~GHz emission is not affected by attenuation. The Balmer decrements measurements by \citet{BLANC09} provided lower values of \Ahalpha{} in the very center of the galaxy. This might reflect to the breakdown of the Balmer decrement at high values of \Ahalpha\ or issues with our aperture photometry in such a crowded, bright region.}
     \label{fig:CompareOtherMeasures_5194}
 \end{figure}

We compare our \pabeta{}-based attenuation measurements to attenuation estimates based on Pa$\alpha$ measurements from \citet{KENNICUTT07}, Balmer decrement measurements from \citet{BLANC09}, and 33~GHz measurements from \citet{QUEREJETA2019} in Fig. \ref{fig:CompareOtherMeasures_5194}. 

For this exercise we use the aperture locations and size (4.3\arcsec in diameter) specified in \citet{BLANC09} to remeasure our \pabeta~maps and the 33~GHz maps. The aperture locations in \citet{BLANC09} cover the inner $4.1 \times 4.1 kpc^2$ of NGC~5194. Here we limit the analysis to only the apertures that also meet the \halpha{} intensity threshold described in Section \ref{sec:analysis}. 

We use local background subtraction as in Section \ref{sec:analysis} to correct for any background variances in the \halpha, \pabeta, and 33~GHz emission images. We find that our results can be very sensitive to the background definition in the center on NGC~5194 due to its large size. To ensure we do not over-subtract the center apertures we use local annuli large enough to sample the empty space around the galaxy center. However, this might bias us to lower background values, and thus we could predict larger luminosities in the galaxy center than is true.

Outside the central 0.3~kpc, our measurements agree well with both the 33~GHz data and the Pa$\alpha$ measurements from \citet{KENNICUTT07} at values of $r > 0.4kpc$. Where $r < 0.4kpc$, we estimate higher values of \Ahalpha.  The disagreement of \pabeta{} corrected \halpha{} and 33~GHz in the galaxy center of NGC~5194 might be caused by our adopted radial gradient converting the 33 GHz data to free free-free emission discussed in Sec. \ref{sec: compare lit}

The Balmer decrement measurements from \citet{BLANC09} are lower than other measurements in the inner $\sim$0.6 kpc of the galaxy. This might imply a mild deviation from a screen geometry, with the Balmer emission being absorbed more easily by higher values of \Ahalpha.

\subsubsection{NGC~6946}

Recombination line measurements of NGC~6946 are not as plentiful as NGC~5194, perhaps due to its low Galactic latitude. 

\textbf{Wide field Br$\gamma$ imaging:} \citet{LI13} mapped NGC~6946 in $Br\gamma$ using the WIRCam (Wide Infrared Camera) on CFHT (Canada–France–Hawaii Telescope). This $Br\gamma$ map was used with the SINGS \halpha{} map to obtain attenuation-corrected \halpha{} values, and thus a SFR. 

Their reported median value of $E(B-V)$ is 0.44~mag. This corresponds to $A_V$ of 1.36 mag assuming $R_V = 3.1$ and a \citet{CCM89}. Our result is slightly higher at $A_V$ = 1.76 mag $\pm$ 0.20 mag.

Part of this discrepancy can be explained by aperture placement. The very center of NGC~6946, where we see our highest values of attenuation, is excluded from the \citet{LI13} analysis. Additionally, our apertures are 2\arcsec{} in diameter. \citet{LI13} use 12\arcsec{} diameter apertures. The larger apertures used by \citet{LI13} might mix low attenuation and high attenuation regions, leading to a lower median attenuation. We also use a different \halpha{} map \citep[we use the one from][while they use the SINGS map]{LONG2019}.

\textbf{Low resolution free-free emission:} \citep{MURPHY11} measured the 33~GHz emission in the nucleus and 9 extra-nuclear regions across NGC~6946. These regions were measured in apertures that were 25\arcsec{} in diameter. They generally found good agreement between the SFR diagnostics and 33~GHz emission. However, in the center the dust-inferred SFR was a factor of 2 larger than that derived using radio data. This was attributed to an accumulation of non-ionizing stars in the center of the galaxy due to an extended episode of star formation. 

\textbf{Attenuation in the nucleus:} We can also compare our median measured attenuation in \halpha~for the nuclear region (which we roughly define here as the central kiloparsec), 2.54 mag, to previous estimates. The attenuation in the nucleus of NGC~6946 has previously been reported as $A_V$ = 4.3 mag \citep[aperture size of 280pc, based on the Balmer decrement][]{ENGELBRACHT96}, $A_V$ = 4.7 mag\citep[90pc aperture size, Balmer decrement][]{QUILLEN01}, and $A_V$ = 5.0 mag \citep[1 kpc, 9.7\micron~silicate absorption feature][]{SMITH07}. Values of $A_V$ have been recorded even higher with attenuation robust IR and Radio data. \citet{TSAI13} measured $A_V \sim 25$ (\Ahalpha$\sim$22) for the central 200 pc using Br$\gamma$ data. Using radio-based gas mass estimates, \citet{SCHINNERER06} estimated the total attenuation in the central $\sim$60pc and found much higher attenuation, as large as $A_V$ $\sim$ 100 mag.

In order to compare our results, we convert our measured values of \Ahalpha{} to $A_V$ using a \citet{CCM89} extinction curve with $R_v = 3.1$. We only have 1 aperture which covers the central 60pc as our aperture size in NGC~6946 is 73~pc. This aperture has a measured $\Ahalpha{}$ of 4.0~mag ($A_V = 4.63$~mag). \citet{ENGELBRACHT96}, \citet{QUILLEN01}, and \citet{SMITH07} found  median $A_V$ attenuation within the central kpc, 280pc, and 90pc to be 2.93 mag, 4.77 mag, and 4.36 mag with a scatter of 1.88 mag, 2.09 mag, and 0.67 mag. Given the differences in aperture definitions and methods, there appears to be good overall agreement between their measurements and ours. We take this as a qualitative confirmation of our measurement, at least at the factor of $\sim 2$ level.

The much larger values of $A_V \sim 100$ ~mag from \citet{SCHINNERER06} likely mainly reflect methodological differences. While our measurements and the Balmer decrement measurements reflect observed recombination line emission, the \citet{SCHINNERER06} estimate expresses the extinction expected based on all of the gas present. The {\sc Hii} regions and gas might not be coincident or the {\sc Hii} regions and gas might be mixed, with some of the {\sc Hii} regions so heavily embedded as to be practically invisible. In either case, geometry can go a long way towards explaining the discrepancy. We return to a similar point comparing emission-based column densities to our absorption measurements in Section \ref{sec:Gas v Ahalpha}.

\section{Results}
\label{sec:results}

We constructed \pabeta{} line emission maps of NGC~5194 and NGC~6946 (Section \ref{sec: making pabeta}). Images of these maps appear in Figure \ref{fig:NGC5194 Mosaic}. 

We compared these \pabeta{} maps to \halpha{} maps from \citet{KENNICUTT07} and \citet{LONG2019} to calculate the implied \halpha\ attenuation and the attenuation-corrected \halpha\ luminosity in 2,075 apertures in NGC~5194 and 1,934 apertures in NGC~6946 (Section \ref{sec:analysis}). These measurements are tabulated in Tables \ref{tab: aperture_table_6946} and \ref{tab: aperture_table_5194}. 

Our maps have wide area, so that the apertures span a large range of galactocentric radius, IR-to-H$\alpha$ color, and gas column density. Note, however, that uncertainties in the background level of the \pabeta\ maps restrict our analysis to regions that are relatively bright in both \pabeta\ and \halpha . Our apertures include 50\% of the total \halpha{} emission (uncorrected for attenuation) from NGC~5194 and 49\% of the total \halpha{} emission from NGC~6946. Although not ideal, this focus on bright regions while subtracting a surrounding ``diffuse'' component follows previous work in the field.

We use these measurements to infer the distribution and radial profile of \Ahalpha\ in each galaxy (Section \ref{sec:AHAdist}). We then test several methods of predicting \Ahalpha\ based on location in the galaxy or measurements at other wavelengths (Section \ref{sec:AHapred}).

\subsection{Distribution of \Ahalpha{}}
\label{sec:AHAdist}

\begin{figure*}[t!]
    \centering
    \includegraphics[width = .45\linewidth]{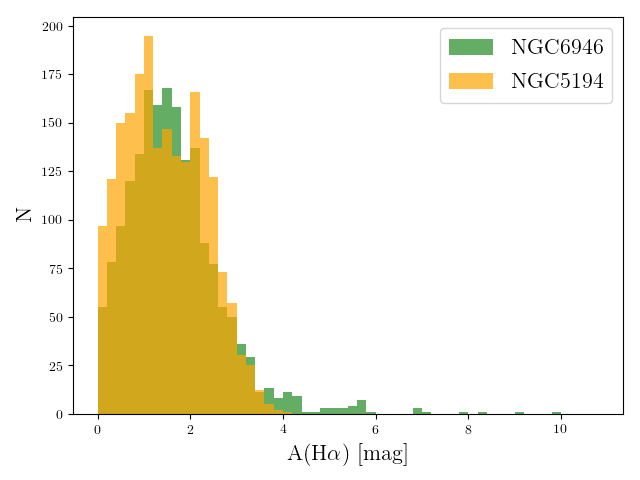}
    \includegraphics[width = .45\linewidth]{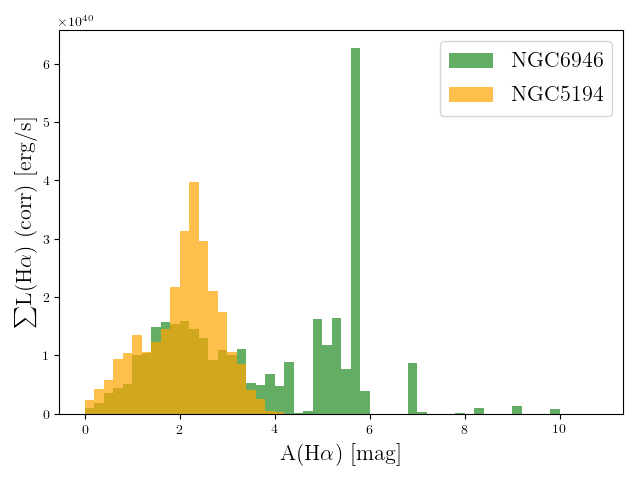}
    \caption{\textbf{Distribution of \Ahalpha{} in apertures across NGC~5194 and NGC~6946} Left: Histogram of the distribution of \Ahalpha{} across NGC 5194 (orange) and NGC 6946 (green), treating each aperture equally. Right: Total attenuation-corrected \halpha{} luminosity contributed by apertures in each bin of \Ahalpha. In both panels, the bins have width of 0.2 mag in \Ahalpha.}
    \label{fig:AHa_Hist_both}
\end{figure*}

\begin{deluxetable}{l|cc}\label{tab:AHaDists}
\tabletypesize{\scriptsize}
\tablecaption{Distribution of \Ahalpha{}.} 
\tablewidth{0pt}
\tablehead{
\colhead{Quantity} &
\colhead{NGC~5194} &
\colhead{NGC~6946} 
}
\startdata
By number\tablenotemark{a} & & \\
... 16$^{\rm th}$ percentile & 0.55~mag & 0.72~mag \\
... median (50$^{\rm th}$ percentile) & 1.41~mag & 1.52~mag \\
... 84$^{\rm th}$ percentile & 2.39~mag & 2.49~mag \\
By luminosity\tablenotemark{b} & & \\
... 16$^{\rm th}$ percentile & 1.15~mag  & 1.62~mag  \\
... median (50$^{\rm th}$ percentile) & 2.19~mag & 3.37~mag \\
... 84$^{\rm th}$ percentile & 2.86~mag  & 5.67~mag \\
\enddata
\tablenotetext{a}{Distribution of \Ahalpha{} treating each aperture as an equal, independent measurement.}
\tablenotetext{b}{Distribution of \Ahalpha{} sorted by the attenuation-corrected \halpha\ luminosity associated with each aperture.}
\tablecomments{Distributions considering measurements from our apertures, which capture $\sim 50\%$ of the \halpha\ emission from each target.}
\end{deluxetable}

Figure \ref{fig:AHa_Hist_both} shows the distribution of \Ahalpha{} in each galaxy. The left panel plots the histogram treating each aperture equally. The right shows the distribution of attenuation-corrected  L(\halpha) as a function of \Ahalpha{}. Table \ref{tab:AHaDists} quantifies both distributions.

Treating all apertures equally, the distribution of \Ahalpha{} appears similar between the two galaxies. Both galaxies show a median \Ahalpha{} near $\sim 1.5$~mag, with $67\%$ of the data lying within roughly $\pm 1$~mag of this median value. These values may be biased somewhat high by our selection of bright apertures, but as we saw above that they agree well with past results. This distribution also appears consistent with measurements by \citet{MURPHY18}. They found a median \Ahalpha{} value of 1.26 $\pm$ 0.09 mag with scatter of 0.87 mag for 162 pointings towards star forming regions across 56 nearby galaxies (with measurements on 30-300 pc scales).  

The histogram of attenuation-corrected \halpha{} luminosity shifts to higher values compared to the histogram treating all apertures equally. That is, the more intrinsically luminous {\sc Hii} regions also tend to be more heavily embedded, with much of the \halpha{} emission from a galaxy occurring in regions of high \Ahalpha{}. We return to this point below.

\begin{figure}[t!]
     \centering
     \includegraphics[width = .95\linewidth]{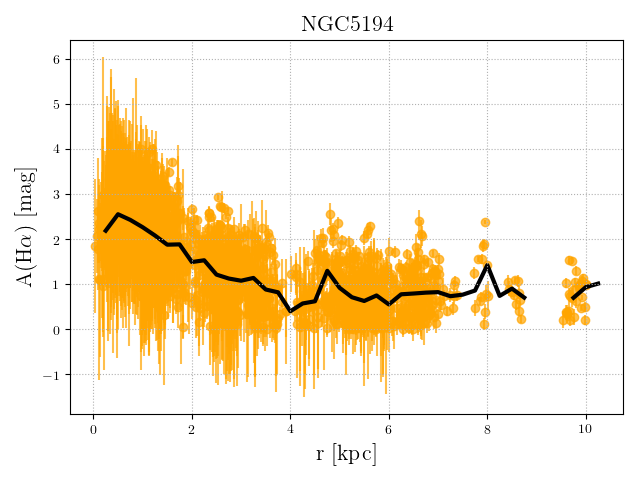}
     \includegraphics[width = .95\linewidth]{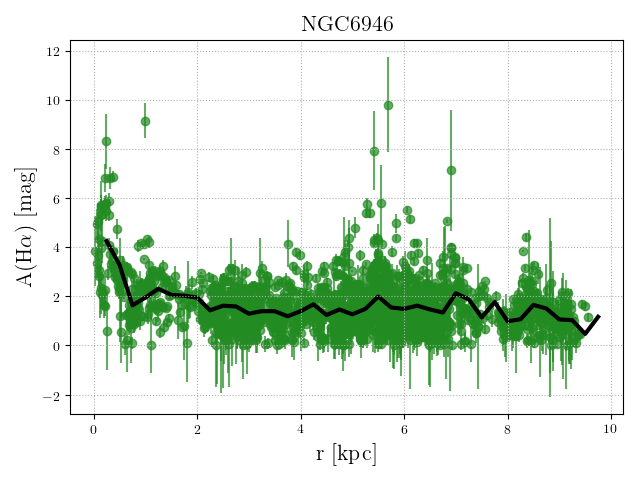}
     \includegraphics[width = 0.95\linewidth]{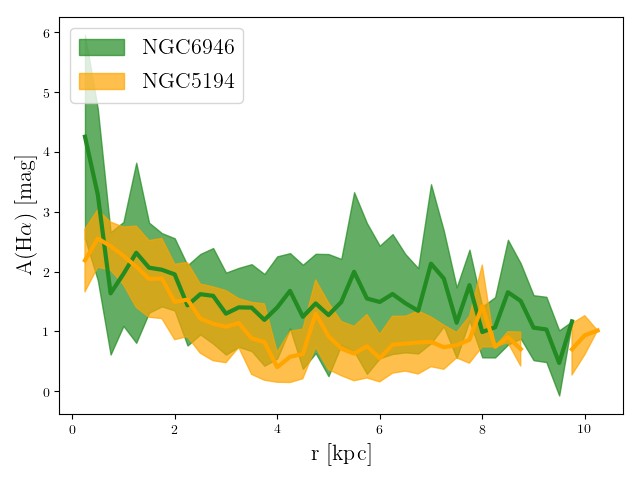}
     \caption{\textbf{\Ahalpha{} as a function of galactocentric radius.} The lines show median \Ahalpha{} in 0.25~kpc-wide bins of galactocentric radius. Points in the top two panels show individual apertures. In the bottom panel the shaded regions show the $\pm 1\sigma$ range in each bin.}
     \label{fig:AharadialDist}
 \end{figure}

Both sets of histograms show structure (e.g., tails and multiple peaks) that corresponds to the morphology of the galaxies. This appears particularly prevalent in the luminosity-weighted histogram. Much of this variation in \Ahalpha{} occurs as a function of radius. We plot this directly, showing \Ahalpha{} as a function of galactocentric radius in Figure \ref{fig:AharadialDist}. 

We observe an overall radial trend in both galaxies. On average, central regions show higher \Ahalpha{} than regions at large radius. NGC~5194 has a large and bright central region, which causes more than 15\% of the apertures to be placed in the center of the galaxy.

This radial trend leads to a bimodal distribution in both histograms in Figure \ref{fig:AHa_Hist_both}. The inner part of the galaxy shows significantly higher attenuation compared to the outer regions. As a result, the two regions appear somewhat distinct in the distributions. The median value of \Ahalpha{} for apertures with galactocentric radius less then 1~kpc is 2.36 mag. By contrast, the  median \Ahalpha{} of apertures with galactocentric radius greater than 1~kpc is 1.14 mag.

The large range of attenuation for NGC~6946 also mostly reflects high attenuation in the galaxy center. The center of NGC~6946 hosts a concentration of molecular gas funneled to the center by an inner stellar bar \citep{SCHINNERER06}. This leads to high gas column densities and correspondingly high attenuations. This gas also forms stars, so that the center of NGC~6946 represents a prime example of a bar-fed nuclear starburst \citep[e.g., see][]{MURPHY11}. 

The very center of NGC~6946 appears more extreme than that of NGC~5194, but the gas also appears more concentrated. While the peak attenuations reach $\gtrsim 4$~mag in the center, they approach the disk-averaged values by $\sim 1$~kpc. By contrast, NGC~5194 shows a more gradual decline in attenuation, with high values out to $\sim 2$~kpc.

Even our high estimates for the central attenuation may underestimate the true values in the innermost part of NGC~6946. The center hardly appears visible in \halpha{} and UV light, so that half of the galaxy center visible in the \pabeta{} image is not visible in the \halpha{} image (see Figure \ref{fig:Contours galaxies}). This might suggest that much of the center might have even higher attenuations than we calculate in our analysis. In fact, \citet{TSAI13} measured $A_V \sim 25$ (\Ahalpha$\sim$22) for the central 200 pc using Br$\gamma$ data. Br$\gamma$ at 2.17\micron, is even more dust robust than \pabeta implying that some of our \pabeta{} emission is likely attenuated before we can observe it. 

A few extranuclear regions in NGC~6946 also show very high attenuation. These are visible in both the maps (Fig. \ref{fig:Aha_spatial}) and radial profiles (Figure \ref{fig:AharadialDist}). While these regions show high \Ahalpha{}, they have only relatively weak actual \halpha{} luminosity. On average, their \halpha{} intensity is only 10\% of the median value of L(\halpha) for all apertures in the galaxy. More, these high \Ahalpha{}, low \halpha{} regions tend to lie near the edge of brighter regions. It is possible that these regions represent a handful of deeply enshrouded, young regions. But given their location and faintness, we suspect that these high \Ahalpha{} mostly reflect instability in the image processing (e.g., local problems with the background subtraction, reprojection, and convolution of the images). Overall, only a few such regions appear in our data, and none are obvious in NGC~5194, so the effect is likely fairly limited.

While these radial gradients give a good first-order picture, both galaxies show significant scatter in \Ahalpha{} at fixed galactocentric radius. In NGC~5194 the typical rms scatter in \Ahalpha{} within a 1~kpc bin is $\approx 0.55$ mag. NGC~6946 shows even a higher variation,  with rms scatter of $\approx 1.0$ mag per 1~kpc bin. In the next sections, we look at how \Ahalpha{} correlates with other quantities.

\begin{deluxetable*}{l|cccccccc}
\tabletypesize{\scriptsize}
\tablecaption{Scatter around predictive models}
\tablewidth{0pt}
\tablehead{
\colhead{} &
\colhead{$\halpha+\alpha\cdot8\micron$} &
\colhead{$\halpha+\alpha\cdot12\micron$} &
\colhead{$\halpha+\alpha\cdot24\micron$} &
\colhead{$\halpha+\alpha\cdot70\micron$} &
\colhead{$\halpha+\alpha\cdot100\micron$} &
\colhead{Gas ($\beta\cdot$Screen)} &
\colhead{Gas (Mixture)} &
\colhead{r} 
}
\startdata
\hline
{}&\multicolumn{8}{c}{NGC~5194}\\
\hline
Linear (mag.) & 0.330 & 0.339 & 0.338 & 0.327 & -- & 0.702 & 0.109 & 0.318\\
Log (dex) & 0.044 & 0.045 & 0.046 & 0.044 &--& 0.099 & 0.218 & 0.105\\
\hline
{}&\multicolumn{8}{c}{NGC~6946}\\
\hline
Linear (mag.) & 0.200 & 0.272 & 0.252 & 0.200 & 0.213 & 0.255 & 0.248 & 0.251\\
Log (dex) & 0.121 & 0.105 & 0.109 & 0.121 & 0.118 & 0.363 & 0.078 & 0.054\\
\hline
{}&\multicolumn{8}{c}{Both Galaxies}\\
\hline
Linear (mag.) & 0.149 & 0.178 & 0.160 & 0.135 &-- &  0.352 & 0.112 & 0.157 \\
Log (dex) & 0.085 & 0.082 & 0.084 & 0.088 & -- & 0.338 & 0.116 & 0.0944
\enddata
\tablecomments{\textbf{Scatter in our data about empirical models to predicts \Ahalpha }. The values of $\alpha$ used in the first four columns can be found in table \ref{tab: IRrelations}. We use our best-fit value for $\alpha$ for all wavelengths. For comparison, the scatter in the data alone, without any model prediction, is 0.84 mag (0.38 dex) in NGC~5194 and 1.04 mag (0.36 dex) in NGC~6946. }
\label{tab:model STDS}
\end{deluxetable*}

\begin{deluxetable}{l|ccc}
\tabletypesize{\scriptsize}
\tablecaption{Correlation coefficients and best fit predictions of \Ahalpha{} and IR bands}
\tablewidth{0pt}
\tablehead{
\colhead{Quantity} &
\colhead{NGC~5194} &
\colhead{NGC~6946} &
\colhead{Both Galaxies}
}
\startdata
{}&\multicolumn{3}{c}{8\micron}\\
\hline
Res. 8\micron & 4\arcsec & 4\arcsec & 4\arcsec\\
r(8\micron) & -0.46 & -0.58 & -0.03\\
$\alpha$(8\micron) & 0.009$\pm$ 0.003 & 0.014 $\pm$ 0.004 & 0.014 $\pm$ 0.004\\
$\alpha$(8\micron) Kenn. 09 & \multicolumn{3}{c}{0.011$\pm$0.003}\\
\hline
{}&\multicolumn{3}{c}{12\micron}\\
\hline
Res. 12\micron & 7.5\arcsec & 7.5\arcsec & 7.5\arcsec\\
r(12\micron) & -0.56 & -0.51 & -0.04\\
$\alpha$(12\micron) & 0.025$\pm$ 0.015 & 0.089$\pm$ 0.027 & 0.051$\pm$ 0.015\\
\hline
{}&\multicolumn{3}{c}{24\micron}\\
\hline
Res. 24\micron & 11\arcsec & 11\arcsec& 11\arcsec\\
r(24\micron) & -0.48 & -0.53 & -0.05 \\
$\alpha$(24\micron) & 0.029$\pm$ 0.009 & 0.049$\pm$ 0.015 & 0.047$\pm$ 0.010\\
$\alpha$(24\micron) Kenn. 07 & \multicolumn{3}{c}{0.038$\pm$0.005}\\
$\alpha$(24\micron) Calz. 07 & \multicolumn{3}{c}{0.031$\pm$ 0.006}\\
\hline
{}&\multicolumn{3}{c}{70\micron}\\
\hline
Res. 70\micron & 8\arcsec & 8\arcsec & 8\arcsec\\
r(70\micron) & -0.59 & -0.55 & -0.04\\
$\alpha$(70\micron) & 0.008$\pm$ 0.003 & 0.014 $\pm$ 0.004 & 0.013$\pm$ 0.004\\
$\alpha$(70\micron) Li 13 & \multicolumn{3}{c}{0.011$\pm$0.001 }\\
\hline
{}&\multicolumn{3}{c}{100\micron}\\
\hline
Res. 100\micron & -- & 9\arcsec & --\\
r(100\micron) & -- & -0.52 & --\\
$\alpha$(100\micron) & -- & 0.018$\pm$ 0.005 & --\\
\enddata
\tablecomments{We tabulate the image resolution, rank correlation coefficients ($r$), and best fit linear coefficient} used to create a linear model.
\label{tab: IRrelations}
\end{deluxetable}

\begin{figure*}[t!]
\centering
\includegraphics[width = 0.45\textwidth]{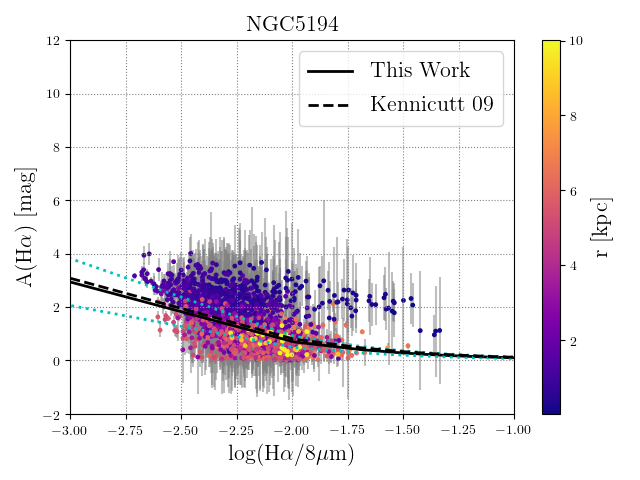} 
\includegraphics[width = 0.45\textwidth]{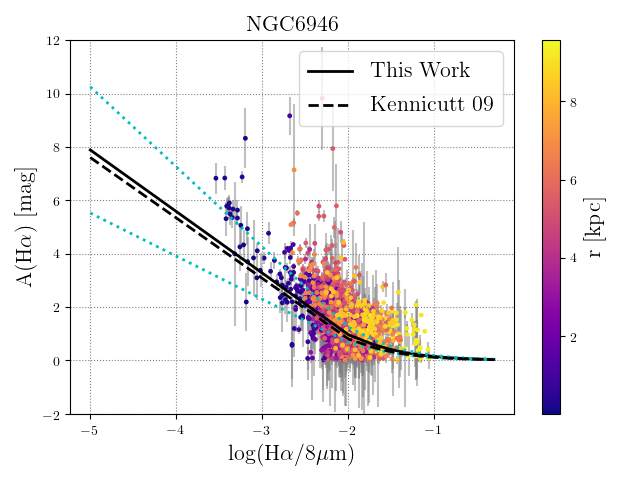} \\
\includegraphics[width = 0.45\textwidth]{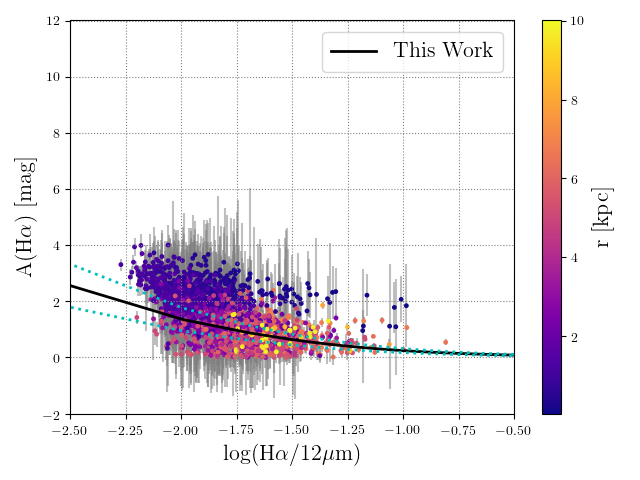}
\includegraphics[width = 0.45\textwidth]{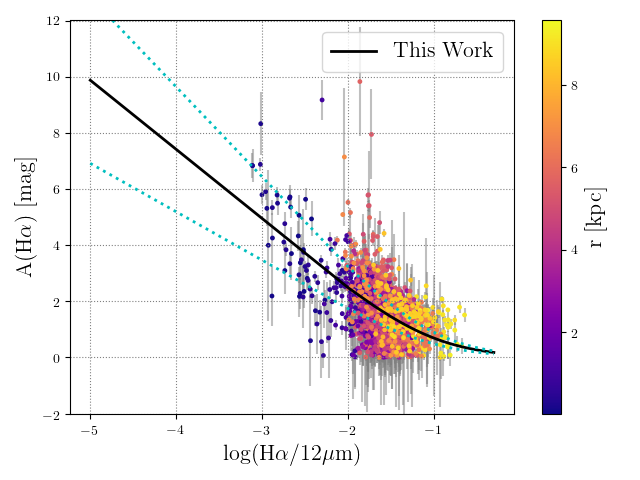} \\
\includegraphics[width = 0.45\textwidth]{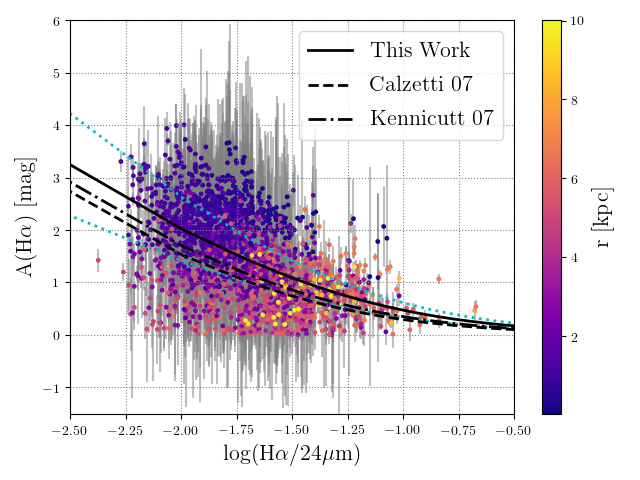}
\includegraphics[width = 0.45\textwidth]{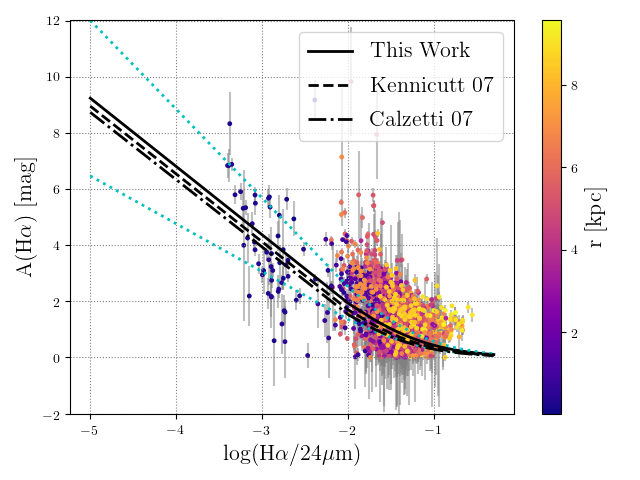} \\
\caption{\textbf{\Ahalpha{} as a function of \halpha{}-to-IR color for mid-infrared bands.} Each panel show \Ahalpha{} as a function of the ratio between \halpha{} emission and IR emission at one band, either 8, 12, or 24\micron. The curves show linear models of the form given in Equation \ref{eq:irmodel}. We plot our best fit $\alpha$ (see Table \ref{tab: IRrelations}) as a solid black curve, with cyan lines indicating $\pm 30\%$ uncertainties on our derived $\alpha$ values. When available, we show literature prescriptions as dashed curves. Note that we derive $\alpha$ excluding the central 2~kpc in NGC~5194 and the center 1~kpc in NGC~6946, which explains some of the low radius deviation from the best-fit trends.} \label{fig:AhavhaIR}
\end{figure*}

\begin{figure}
\centering
\includegraphics[width = 0.95\linewidth]{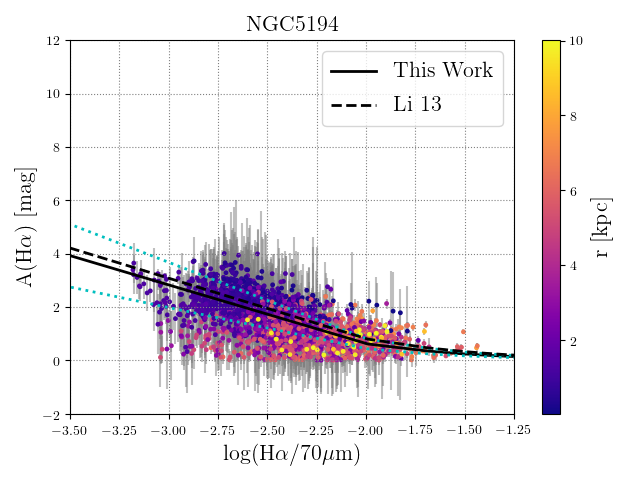}
\includegraphics[width = 0.95\linewidth]{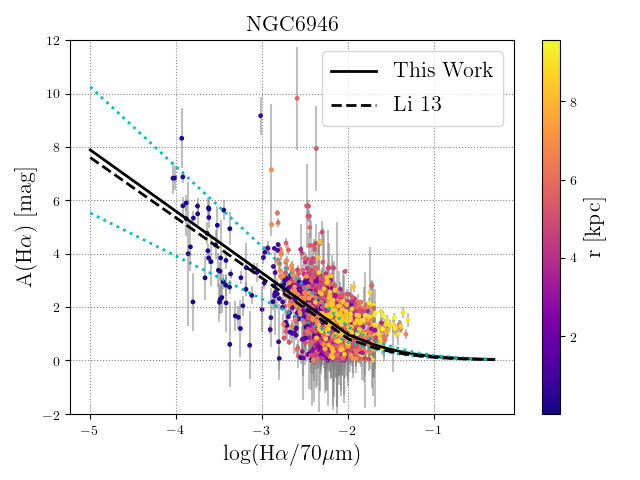}
\includegraphics[width = 0.95\linewidth]{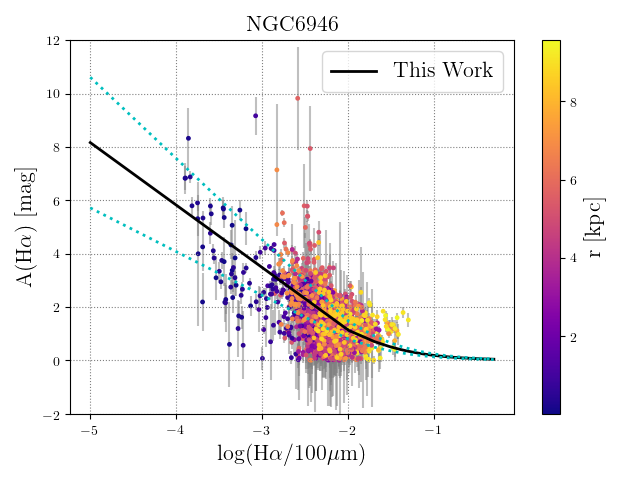}
\caption{\textbf{\Ahalpha{} as a function of \halpha{}-to-IR color for far-infrared bands.} Similar to Figure \ref{fig:AhavhaIR}, each panel shows \Ahalpha{} as a function of the ratio between \halpha{} emission and far infrared emission, either at at 70 or (for NGC~6946) 100\micron. The curves show linear models of the form in Equation \ref{eq:irmodel}. We show both our best fit $\alpha$ (solid black curves) and literature prescriptions (dashed curves). These are summarized in Table \ref{tab: IRrelations}. Cyan lines indicate $\pm 30\%$ uncertainties on our derived $\alpha$ values. Our fit to $\alpha$ excludes the central 2~kpc in NGC~5194 and the center 1~kpc in NGC~6946. The apertures are color coded by radius in kpc.}
\label{fig:AhavhaIR_FIR}
\end{figure}

\subsection{Relation between \Ahalpha{} and other observable quantities} 
\label{sec:AHapred}

Mapping Paschen $\beta$ across two of the brightest, nearest star-forming galaxies required a significant time investment from \textit{Hubble}. Given the difficulty in obtaining these ``gold standard'' estimates of the ionizing photon rate, it has become standard practice to predict \Ahalpha{} via other less direct means (see Section \ref{sec:intro}). A main motivation for obtaining and analyzing these maps was to further evaluate common empirical prescriptions for \Ahalpha{}. 

In this section, we compare our \Ahalpha{} estimates to the local \halpha{}-to-infrared color, the \halpha{} luminosity of the region, galactocentric radius, and the local gas column density. We evaluate the ability of each quantity to predict \Ahalpha{} in one or both galaxies. We measure the scatter, and calibrate free parameters in the empirical prescriptions. We tabulate these results in table \ref{tab:model STDS}. 

\subsubsection{\Ahalpha{} from \halpha{} and infrared emission}
\label{sec:Halpha+IR}

\citet{CALZETTI07} and \citet{KENNICUTT07} showed that \Ahalpha{} can be estimated from the ratio of infrared to \halpha\ emission. Where there is more IR emission relative to \halpha{}, the \halpha{} line tends to suffer from more attenuation. Because IR measurements have become plentiful thanks to \spitzer, \herschel, and WISE, this comparison to IR emission now represents a main way to estimate \Ahalpha{}.

As discussed in Section \ref{sec:data}, we do not consider the background level of our \pabeta{} maps stable enough to measure a robust intensity at the low resolution of our IR data (4, 7.5, 11, 8, and 9\arcsec{} for the 8, 12, 24, 70, and 100\micron~bands respectively). There is simply too much uncertain, low-level background structure involved in the convolution to yield a robust result.

To work around this limitation, we compare \Ahalpha{} estimated from combining \pabeta{} and \halpha{} within our $2''$ apertures to the \halpha -to-IR ratio measured at lower resolution. We convolve the archival \halpha{} images to match the PSF of each IR band. We then divide the \halpha{} image by the IR bands to create an \halpha-to-IR ratio image for each IR band at the lower IR resolution. We sample these lower resolution images using the same aperture locations and sizes used to create our high resolution attenuation maps (see Section \ref{sec:analysis}).

This approach yields low resolution \halpha{}-to-IR colors, which we compare to high resolution attenuation estimates. This situation actually resembles a common application of these prescriptions. The PSF of IR telescopes is almost always poor compared to that of seeing-limited optical images. The case that we consider resembles using low-resolution IR data to attempt to correct a high-resolution \halpha{} map for attenuation.

Before sampling we verify that the median values of apertures placed on empty sky was indeed 0 erg~s$^{-1}$~cm$^{-2}$ (or not a number if the IR band in the denominator had a $\sim 0$ background). We do not conduct local background subtraction on any measurements used to calculate the  \halpha-to-IR ratio. Instead we simply take the sum of the aperture and divide that by the aperture area to obtain an average IR value. This might bias us somewhat high in IR, which would bias our \halpha-to-IR value low. We tested this by applying apertures the size of the PSF to the image. We compared the results we have with a PSF sized aperture to those calculated with our adopted 2\arcsec{} aperture. The median difference between the \halpha-to-IR ratio  to the \halpha-to-IR  ratio  was 0.15 mag with a scatter of 0.09 mag. This indicates that we may be underestimating our \halpha-to-IR ratio, but only slightly.

We compare this lower-resolution \halpha{}-to-infrared ratio to our \pabeta{}-based \Ahalpha{} estimates in Figures \ref{fig:AhavhaIR} and \ref{fig:AhavhaIR_FIR}. Table \ref{tab: IRrelations} reports rank correlation coefficients relating the \halpha{}-to-IR color to \Ahalpha{} for each band. 

Both the Figures and the table show that for all IR bands, the ratio of \halpha{}-to-IR anti-correlates with \Ahalpha{}. That is, as \halpha{}-to-IR goes up, \Ahalpha{} goes down and vice versa. Typical rank correlation coefficients lie in the range $-0.45$ to $-0.6$. This is exactly the sense of the \citet{CALZETTI07} and \citet{KENNICUTT07} relations, and this clearly also holds in our data.

As in \citet{CALZETTI07} and \citet{KENNICUTT07}, we calibrate linear relations between \halpha{}, IR, and attenuation-corrected \Ahalpha{}. This linear approach has the advantage of being robust when used across many different spatial scales, at the cost of some potential inaccuracy. We recast this into a model relating \Ahalpha{} to the \halpha{}-to-IR ratio:

\begin{eqnarray}
\label{eq:irmodel}
L(\halpha{})_{\rm corr} &=& L(\halpha{}) + \alpha \times \left( \nu L_\nu (IR) \right) \\
L(\halpha{})_{\rm corr} &=& 10^{\Ahalpha{}/2.5} \times L(\halpha{}) \\
\Ahalpha{} &=& 2.5 \log_{10} \left( 1 + \alpha \times \frac{\nu L_\nu (IR)}{L_{\halpha{}}} \right)
\end{eqnarray}

\noindent where $L(\halpha)_{\rm corr}$ is the attenuation-corrected \halpha{} luminosity in a region, $L(\halpha)$ is the measured (i.e., not attenuation-corrected) luminosity in a region, $\alpha$ is an empirical scaling coefficient, and $\nu L_\nu$ is the infrared luminosity of a region. In our cross-scale approach \Ahalpha{} is measured at high resolution and $\nu L_\nu / L(\halpha{})$ is measured at low, matched resolution.

For each IR band, we calculate the correlation coefficients and the empirical scaling factor, $\alpha$, to translate the \halpha-to-IR ratio into an estimate of attenuation. We report these for each galaxy and for both galaxies combined in Table \ref{tab: IRrelations}. In Table \ref{tab:model STDS}, we also give the scatter in \Ahalpha{} about the predictions from each prescription for \Ahalpha{}. Wherever a literature \halpha-to-IR ratio is available we plot the expected relation. These are tabulated in Table \ref{tab: IRrelations}.

In both galaxies the \halpha-to-70\micron{} ratio predicts \Ahalpha{} with the least amount of scatter. It also boasts the highest correlation coefficient in NGC~5194 and the second highest in NGC~6946. As shown in Table \ref{tab:model STDS}, the \halpha-to-70\micron{} not only produces the least amount of scatter in \Ahalpha{} among the IR models, but produces the least amount of scatter in \Ahalpha{} overall. For galaxies like NGC~5194 and NGC~6946,  we find the \halpha-to-70\micron{} ratio best to predict \Ahalpha.

\textbf{Central regions:} When we calculate the factor $\alpha$, we omit the central kiloparsec in NGC~6946 and the central 2 kiloparsecs in NGC~5194. This ensures that we are not biased by the AGN in NGC~5194 and ensures that we only use apertures where we are confident in the background subtraction.

Figure \ref{fig:AhavhaIR} shows that this choice to exclude the centers does matter. While a single linear coefficient has trouble predicting \Ahalpha{} for all points, our measurements at larger radii, which lie mostly in the spiral arms, agree reasonably with previous results. 

This deviation of the center from the overall relationship may reflect a different infrared SED and dust geometry in the galaxy center compared to the disk. In NGC~5194, it might also partially reflect the influence of the AGN. Alternatively, this may reflect the difficulties in establishing a good zero-point for the bright, extended central regions, as discussed above.

In NGC~6946, the very center of the galaxy clearly behaves separately from the rest of the galaxy. Points that have very high values of \Ahalpha{} also have very low values of \halpha-to-IR. As these points are very bright in the IR, this indicates that the high values of \Ahalpha{} calculated have merit. Additionally the high values of \Ahalpha{} fall along the predicted trends from previous literature measurements. However, the apertures that show very high attenuation values at larger radii from the galaxy center seem to be simple Poisson differences between the \halpha{} map and our \pabeta{} map as points do not seem as bright in the IR as we would expect. 

\subsubsection{\Ahalpha{} from \halpha{} luminosity of each region}

\begin{figure*}
    \centering
    \includegraphics[width = 0.45\textwidth]{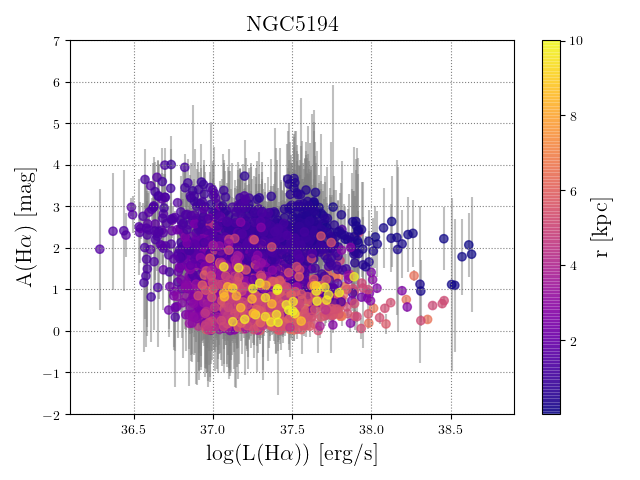}    
    \includegraphics[width = 0.45\textwidth]{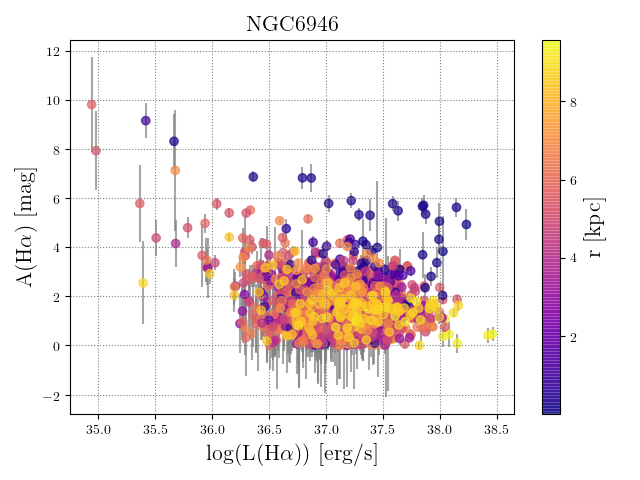}
    \includegraphics[width = 0.45\textwidth]{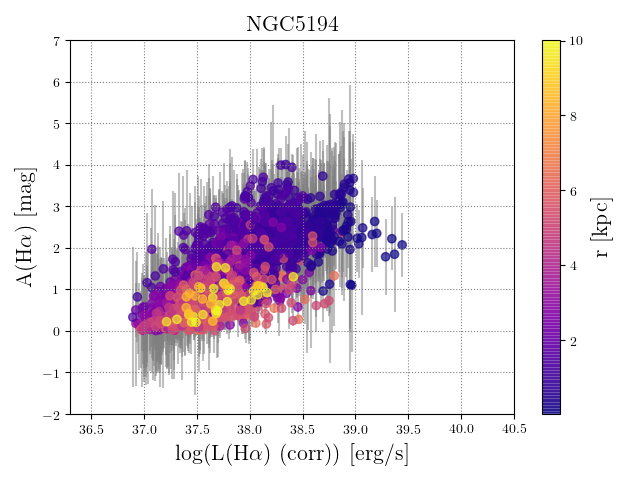}
    \includegraphics[width = 0.45\textwidth]{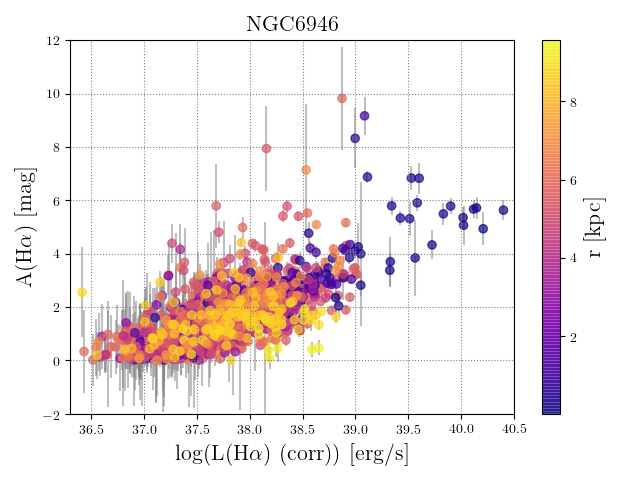}
    \caption{\textbf{Comparison of \Ahalpha{} to \halpha{} luminosity.} In the top panels we compare the attenuation in \halpha{} to the luminosity in the same aperture \textit{before} we correct for attenuation. NGC~5194 is on the left and NGC~6946 is on the right. We find no obvious correlation between the two. In the bottom panels, we compare the attenuation in \halpha{} to the luminosity in the same aperture \textit{after} correcting for attenuation. There is a strong positive correlation between \Ahalpha\ and attenuation-corrected luminosity.}
    \label{fig:LuminosityPlots}
\end{figure*}

In Figure \ref{fig:LuminosityPlots}, we compare the calculated value of \Ahalpha\ to the \halpha{} luminosity. We plot both the observed luminosity and the attenuation-corrected value.

Comparing \Ahalpha{} to the observed \halpha{} luminosity tests whether regions that appear bright before any attenuation correction tend to have high \Ahalpha{}. We find no significant correlation between L(\halpha) and \Ahalpha. The correlation coefficient between the two is -0.06 for NGC~5194 and -0.21 for NGC~6946. Apertures with large values of \Ahalpha{} do not appear preferentially bright or dim. 

The situation appears dramatically different when we compare \Ahalpha{} to the attenuation-corrected \halpha{} luminosity. Here we find a strong correlation. NGC~5194 shows a rank correlation coefficient of 0.71 between \Ahalpha{} and attenuation-corrected \halpha{} luminosity, and NGC~6946 shows a rank correlation coefficient of 0.68. The $p$-values for both are effectively zero. Some of this correlation will be a artifact. \Ahalpha{} affects both axes, so that noise in \Ahalpha{} induces a positive correlation. But we also saw in the previous section that \Ahalpha{} appears higher in IR-bright regions and the inner galaxy. These are systematic trends, not the result of noise, and they have the same sense as what we observe here. More heavily embedded regions, which will also tend to be more IR-bright and lie at lower galactocentric radius, appear more heavily extinguished. And as Figure \ref{fig:LuminosityPlots} shows, they also show higher total attenuation-corrected luminosity.

Summarizing, Figure \ref{fig:LuminosityPlots} shows that the most intrinsically luminous apertures do appear to show the highest attenuation. This also appears consistent with the previous two section. Put another way, more intrinsically bright {\sc Hii} region are more likely to be heavily affected by dust. However, this conclusion only applies to the attenuation-corrected \halpha{} luminosity (bottom panels). The \textit{observed} \halpha{} luminosity (top panels) does not show a similar trend, due to the effects of attenuation. Therefore we do not recommend using observed \halpha{} luminosity to predict \Ahalpha.

\subsubsection{\Ahalpha from galactocentric radius}

Figure \ref{fig:AharadialDist} shows that \Ahalpha{} does correlate with galactocentric radius.  In both galaxies, we find higher \Ahalpha{} near the galaxy center and lower values at large radii. But the bottom panel of Figure \ref{fig:AharadialDist} also shows that the radial distribution of \Ahalpha{} differs between NGC~5194 and NGC~6946. Simply knowing the galactocentric radius, without know what galaxy one is looking at does not offer a good general predictor of \Ahalpha{}.

To quantify this, we fit an exponential profile to \Ahalpha{} as a function of radius in each galaxy. While the exponential can predict the attenuation in \Ahalpha{} well for each galaxy individually, the values of the parameters in the fits vary significantly from NGC~6946 ($b=1.876$) to NGC~5194 ($b=0.491$), to both galaxies together ($b=0.991$). 

If we had a larger sample size we might be able to fit a general profile, e.g., in which the amplitude and scale length of the \Ahalpha{} profile depended on galaxy properties. With a sample size of two, we are not in a position to do more than comment that the profiles appear different. \Ahalpha{} drops with radius in both cases, but the profile has different scale lengths and amplitudes in the two galaxies.

\subsubsection{\Ahalpha from gas column density}
\label{sec:Gas v Ahalpha}

The dust that causes attenuation will be mixed with gas, which is mostly either atomic, {\sc Hi}, or molecular, H$_2$. Thus, the column density of gas should be related to the observed attenuation, modulo uncertainties regarding the geometry and dust-to-gas ratio. The {\sc Hi} gas column can be traced by 21-cm line emission, while the H$_2$ gas column can be traced by CO line emission.

\begin{figure*}
    \centering
    \includegraphics[width = .45\linewidth]{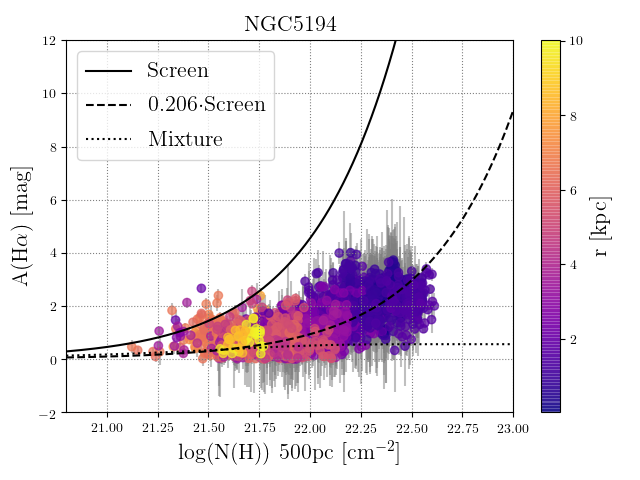}
    \includegraphics[width = .45\linewidth]{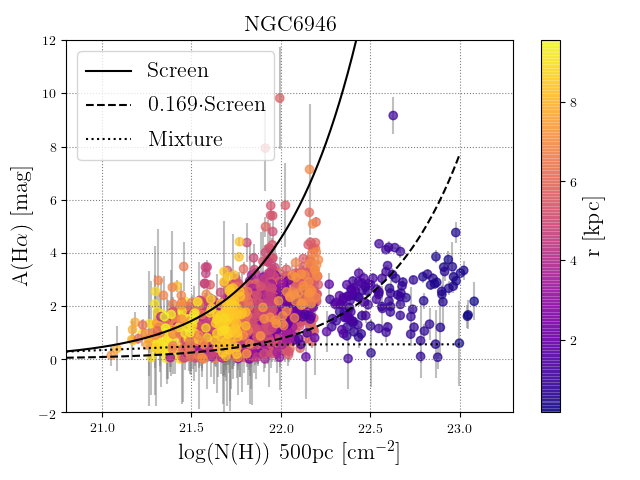}
    \includegraphics[width = .45\linewidth]{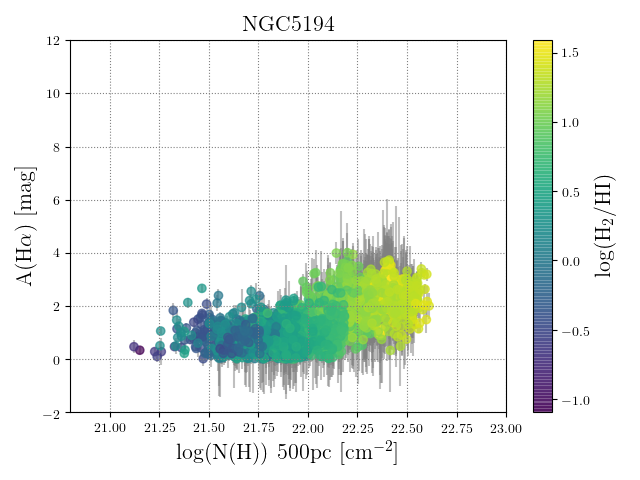}
    \includegraphics[width = .45\linewidth]{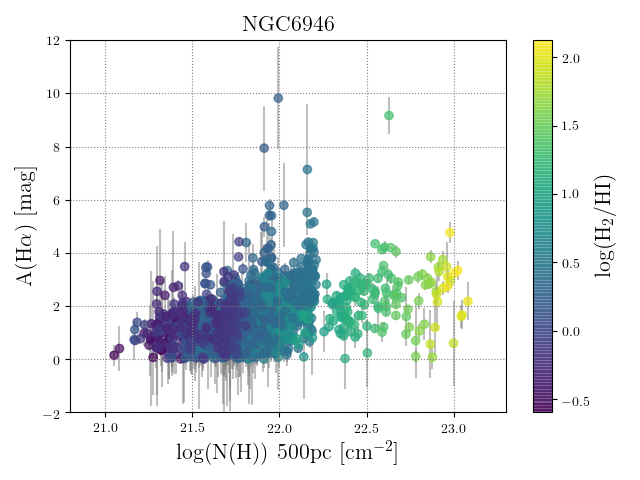}
    \includegraphics[width = .45\linewidth]{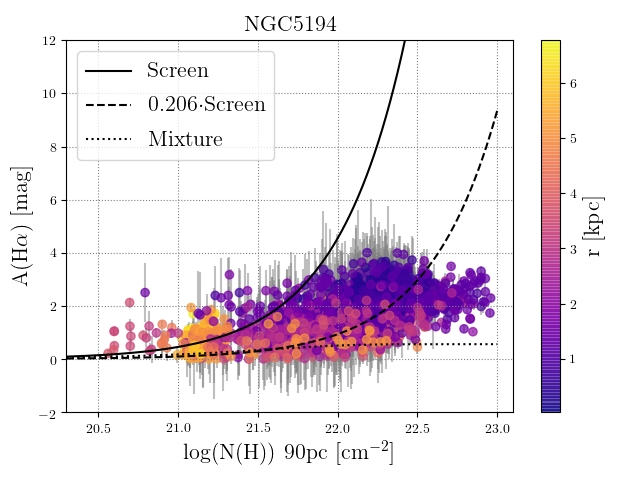}
    \includegraphics[width = .45\linewidth]{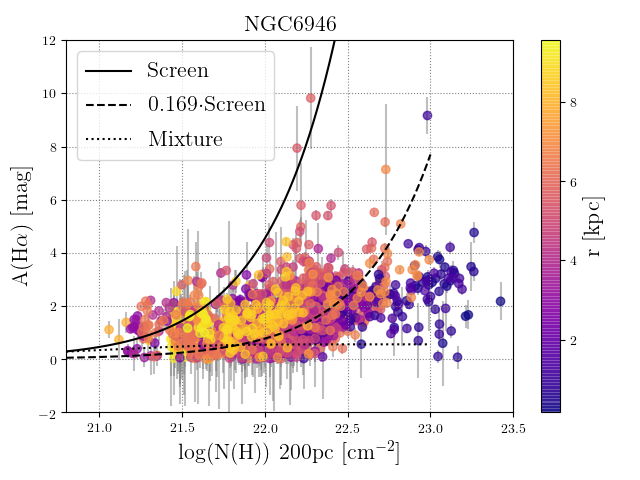}
    \caption{\textbf{\Ahalpha{} as a function of total ({\sc Hi}+H$_2$) column density.} Our calculated \Ahalpha{} as a function of atomic and molecular gas column density estimated from radio emission line maps. Top: Left: NGC~5194 shows a strong correlation between \Ahalpha{} and $N(H)$, which matches a screen model reasonably. Right: NGC~6946 also shows a correlation between \Ahalpha{} and $N(H)$, but with a bimodal distribution, similar to the behaviour shown in Figure \ref{fig:AhavhaIR}. A screen model where a quarter of the gas contributes to attenuation fits most of the NGC~6946 data well and fits the NGC~5194 data very well. For comparison, we also plot a mixture model (dashed gray line) in which \halpha\ and gas are evenly mixed. In the top and bottom rows, point are color coded by galactocentric radius. In the middle row, points are color coded by the ratio of $H_2$ to {\sc Hi} column in the aperture. In both galaxies, low \Ahalpha{} maps to low column, high radius, and low $H_2$/{\sc Hi}. The bottom row uses the highest resolution molecular gas map available ($\sim 90$~pc in NGC~5194 and $\sim 200$~pc in 6946). The higher resolution map leads to a more continuous behavior at low radii in NGC~6946, though CO-to-H$_2$ conversion fact effects may also explain the anomalous behavior of the central kpc.}
    \label{fig: GasAHa_radii}
\end{figure*}

\begin{figure}
\centering
\includegraphics[width=.95\linewidth]{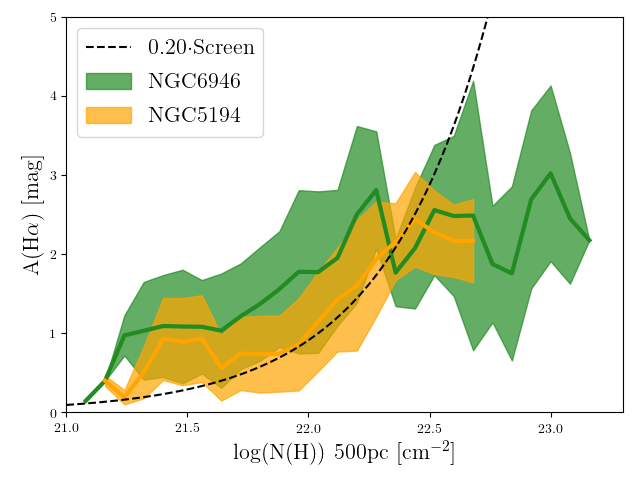}
\caption{\textbf{\Ahalpha{} vs. $N(H)$ for both galaxies.} Binned trend in \Ahalpha{} as a function of total H column at 500~pc resolution for both galaxies. Lines show the median \Ahalpha{} at fixed $N(H)$ and the shaded region shows the $\pm 1\sigma$ scatter region. The two galaxies agree relatively well at intermediate $N(H)$. Both appear roughly consistent with a foreground screen with magnitude $\gamma = 0.2$.}
\label{fig: GasAHa_both}
\end{figure}

In Figures \ref{fig: GasAHa_radii} and \ref{fig: GasAHa_both} we plot \Ahalpha{} as a function of neutral gas column density. In all three Figures, the $x$-axis shows the total column density of neutral hydrogen, N(H), calculated via:

\begin{equation}
N(H) = N({\rm HI}) + 2\cdot N(H_2)~.
\end{equation}

\noindent We calculated $N({\rm {\sc HI}})$ from 21-cm line maps and $N(H_2)$ from CO line maps as described in Section \ref{sec:column}. In the top panels of Figures \ref{fig: GasAHa_radii} both the CO and {\sc Hi} data have been convolved to a matched $13.3'' \sim 500$~pc resolution. In the bottom panels, we use higher resolution CO data combined with the native (but still low) resolution {\sc Hi}. These CO data have resolution $\sim 90$~pc for NGC~5194 and $\sim 200$~pc for NGC~6946. In both cases we assume that the {\sc Hi}, which has coarser resolution, forms a smooth background. In comparing a low resolution mean column density to a more local \Ahalpha{}, this exercise resembles what we did for the IR-to-\halpha{} color.

We tabulate the correlation coefficients between \Ahalpha{} and gas column density in Table \ref{tab:GasDists}. In Figure \ref{fig: GasAHa_radii} we plot \Ahalpha{} as a function of $N(H)$ for both galaxies. The top four panels show results at $13''.3 \sim 500$~pc resolution. The bottom panels show results at the highest available resolution. Color coding in the panels reflects both galactocentric radius and the molecular-to-atomic gas ratio.

In each case, we observe a significant correlation, such that we find higher \Ahalpha{} at higher $N(H)$.  The rank correlation coefficients lie in the range 0.4-0.7, comparable to what we find for the IR-to-\halpha{} ratio. That is, \Ahalpha{} correlates with the local gas column density about as well as it correlates with the IR-to-\halpha{} color. For NGC~6946 the correlation improves somewhat, from $0.42$ to $0.47$, if we use the high resolution CO instead of the $13.3''$ data. The Figure shows that this partially reflects that with higher resolution, the bright, high column central region forms a more continuous distribution with the disk.

Figure \ref{fig: GasAHa_radii} also shows that, as expected, gas columns are higher at low radii and that the ISM is mostly molecular in most of our apertures. The molecular fraction correlates with the total column density such the the high-column and low-radius points tend to be overwhelmingly molecular.

The center of NGC~6946 stands out in these plots, with high column density compared to its \Ahalpha{}. This likely at least partially reflects variations in the CO-to-H$_2$ conversion factor. We adopted a fixed CO-to-H$_2$ conversion factor, but \citet{SANDSTROM13} have shown that $\alpha_{\rm CO}$ shows a strong radial gradient in this galaxy, with much lower values near the galaxy center. Applying such a variable conversion factor will move the central points to lower column density without affecting their \Ahalpha{}, bringing the center into better agreement with the rest of the galaxy.

We compare the galaxies to each other in Figure \ref{fig: GasAHa_both}. The two galaxies roughly similar trends, with higher values of \Ahalpha{} indicating higher values of gas column density. In detail, NGC~5194 shows a better-behaved relation whereas NGC~6946 shows more scatter in both column density and attenuation. 

\begin{deluxetable}{l|cc}\label{tab:GasDists}
\tabletypesize{\scriptsize}
\tablecaption{Comparing $N(H)$ and \Ahalpha.} 
\tablewidth{0pt}
\tablehead{
\colhead{Quantity} &
\colhead{NGC~5194} &
\colhead{NGC~6946}
}
\startdata
$r$ for $N(H)$ at 500~pc\tablenotemark{a} & 0.67 & 0.42\\
$r$ for $N(H)$ at 90~pc & 0.60 & -- \\
$r$ for $N(H)$ at 200~pc & --& 0.47\\
Screen $\gamma$\tablenotemark{b} & 0.206 & 0.169 \\
Screen $\gamma$ --- both galaxies & \multicolumn{2}{c}{0.190} \\
\enddata
\tablenotetext{a}{Rank correlation coefficient relating \Ahalpha{} to total $N(H) = N({\rm HI})+2\cdot N(H_2)$.}
\tablenotetext{b}{Best-fit conversion from column to \Ahalpha{} relative to the expectation for a foreground screen with Milky Way composition (Equations \ref{eq: AHa gas relation} and \ref{eq:gamma}).}
\end{deluxetable}

\textbf{Comparison to simple models:} Quantitatively, the relationship between \Ahalpha{} and $N(H)$ will reflect the dust-to-gas ratio and the relative placement of dust, \halpha{} emission, and the observer. To explore this, we  compare our observations to two simple geometric models, a foreground screen of dust and a smooth mixture of \halpha{} emission and dust. 

In both models, we allow the overall normalization of the model to vary. For a Milky Way dust-to-gas ratio, all of the gas column in a foreground screen, and a \citet{CCM89} extinction curve, we expect

\begin{equation}\label{eq: AHa gas relation}
    A(H\alpha)^{\rm MW} \sim 0.86 \cdot \frac{N_H}{1.9\times10^{21} {\rm cm}^{-2}}~{\rm mag}~.
\end{equation}

\noindent Deviations from this normalization could reflect that some of the gas is behind the emitting source or that the gas and \halpha{} emission have different distributions on small scales. Most simply, we might expect 50\% of the gas to lie behind the emission in a case where \halpha{} comes from a thin layer at the midplane.

We calculate the best-fit screen and mixture models. Here, a ``mixture model'' means that instead of all light passing through a uniform screen of dust with some fixed value of A(H$\alpha$), we consider the \halpha{} emission and attenuation to be evenly mixed. Thus, some of the \halpha{} emission experiences only low \Ahalpha{}, while other \halpha{} emission experiences attenuation corresponding to the full column density of dust. We implement this model numerically, evenly distributing the \halpha{} emission across \Ahalpha{} up to the maximum value of \Ahalpha{}.

For the screen models, we scale Equation \ref{eq: AHa gas relation} by a factor $\gamma$ so that:

\begin{equation}
\label{eq:gamma}
\gamma \equiv \frac{\Ahalpha{}}{\Ahalpha{}^{\rm MW}}
\end{equation}

\noindent with $\Ahalpha{}^{\rm MW}$ from Equation \ref{eq: AHa gas relation}. In the mixture models, we perform an analogous calculation, scaling the total column through the screen by $\gamma$.

Figure \ref{fig: GasAHa_radii} shows the curves for a full screen of Milky Way-style gas, the best-fit fractional screen (i.e., the full Milky Way screen scaled by the best-fit $\gamma$), and the best-fit mixture model. In both galaxies, the full screen model overpredicts the amount of attenuation. Instead, the best-fit screens have $\gamma=0.206$ in NGC~5194, $\gamma=0.169$ in NGC~6946, and $\gamma = 0.19$ combining both galaxies. This fractional screen model works particularly well in NGC~5194. In NGC~6946, the fractional screen can describe the data at large radius well, but fails to simultaneously capture the galactic center.

We find $\gamma \sim 0.2$. We would expect $\gamma \sim 0.5$ from a smooth screen with half of the gas on the near side of the \halpha{}-emitting region. In principle, lower $\gamma$ could result from a lower dust-to-gas ratio, but neither NGC~5194 and NGC~6946 are notably low metallicity or dust-poor \citep[e.g., see][]{ANIANO19}. Another, more likely possibility, is that the gas and dust are patchy, with \halpha{} emission and gas imperfectly aligned at small scales. Observations comparing \halpha{} and CO emission at $\sim 1{-}3''$ resolution, including in NGC~5194, appear to show exactly this. \citet{SCHINNERER17} observed offsets between star formation tracers along a spiral arm in NGC~5194. In a systematic analysis of eight spiral galaxies including NGC~5194, \citet[][]{SCHINNERER19} show directly only $\sim 50\%$ of bright \halpha{} and CO are coincident  at 140~pc resolution. This is very similar to the scales on which we conduct our analysis and these two results appear consistent.

In this case $\gamma \sim 0.2$ could reflect that half of the dust lies behind the emission, implying $\gamma \sim 0.5$, and that only $\sim 40\%$ of the dust in front of the \halpha{} actually spatially overlaps the \halpha{}. Only this overlapping gas will lead to attenuation, so the effective $\gamma \sim 0.5 \times 0.4 = 0.2$. This patchy scenario would still be consistent with the agreement between the Balmer decrement, \pabeta , and 33~GHz emission. It only implies that some of the gas is not actively involved in producing attenuation. We do caution that the resolution mismatch between the gas and the \Ahalpha{} estimate renders these numbers even more approximate.

After we fit one free parameter, the gas performs about as well as the IR emission in predicting \Ahalpha{}. This could be expected, since rank correlation coefficient relating \Ahalpha{} to $N(H)$ resembles that seen between \Ahalpha{} and IR-to-\halpha{}.

This does not necessarily make gas an equally good predictor of \Ahalpha{} compared to IR emission. Some IR bands can be used to predict \Ahalpha{} in a way that is robust to metallicity effects and remarkably stable across systems \citep[e.g., see][]{CALZETTI07,CALZETTI10}. On the other hand, the gas-to-dust ratio has been shown to depend on metallicity \citep[][]{DRAINE07,REMYRUYER14,ANIANO19} and the geometry for the gas remains a major uncertainty. We do not necessarily expect $\gamma = 0.2$ to hold generally in the same way as some of the IR-based $\alpha$ values. Still, these kind of gas-attenuation comparisons offer the prospect to learn more about the relative geometry of \halpha{} and dust and could be calibrated into a general tool in future work.

\section{Summary and conclusions}

We utilized new HST WFC3 \pabeta{} observations of NGC~6946 and archival HST \pabeta{} observations of NGC~5194 to create galaxy-wide \pabeta{} maps. We compared these with archival \halpha{} images to estimate the attenuation affecting the \halpha{} line, \Ahalpha{}. 

We estimate \Ahalpha\ in 2\arcsec{} diameter apertures placed to cover all regions where \halpha{} emission appears bright enough that we expect to detect \pabeta . These apertures cover $\sim 50\%$ of the overall \halpha{} flux from each galaxy. Because the WFC3 mosaics have wide areal coverage, we are able to place many apertures spanning a large range of galactocentric radius in both galaxies. In total, we measure fluxes and line ratios for 2,075 apertures in NGC~5194 and 1,934 apertures in NGC~6946. We tabulate these measurements, the calculated \Ahalpha{}, and several other quantities of interest in Tables \ref{tab: aperture_table_5194} and \ref{tab: aperture_table_6946}. In NGC~5194 we compare our measurements to recent work using 33~GHz radio data to trace free-free-emission \citep{QUEREJETA2019}, finding overall consistency between the two data sets.

Treating all apertures equally, NGC~5194 has median \Ahalpha{} of 1.4~mag with a 16-84$^{\rm th}$ percentile range of 0.6-2.4~mag. NGC~6946 has median  \Ahalpha{} of 1.5~mag with 16-84$^{\rm th}$ percentile range 0.7-2.5. The luminosity-weighted mean \Ahalpha{} is higher in both galaxies, $\sim 2.2$~mag in NGC~5194 and $\sim 3.4$~mag in NGC~6946. In both cases, this reflects heavy attenuation in the inner regions of the galaxy. 

In general, we find that the most intrinsically luminous regions show the highest \Ahalpha{}. However this high attenuation diminishes the apparent \halpha{} luminosity of these regions, so that observed \halpha{} luminosity on its own is not a good predictor of \Ahalpha . Instead, galactocentric radius, the IR-to-\halpha{} ratio, and the local gas column density can all be used to predict \Ahalpha{} with varying degrees of accuracy.

We compare \Ahalpha\ to the local IR-to-\halpha{} ratio, using IR emission measured at 8, 12, 24, 70, and (for NGC~6946) 100$\mu$m. Due to limitations in the data, we compare our $2''$ resolution \Ahalpha{} estimates to the IR-to-\halpha{} ratio measured at lower resolution. This resembles the real practical case where IR emission is only available at low resolution and one might wish to estimate the higher resolution, e.g., seeing limited, \Ahalpha{}. We find a strong anti-correlation between the \Ahalpha{} and the IR-to-\halpha{}, in good agreement with \citet{CALZETTI07}, \cite{KENNICUTT07}, and much following work. For each band, each galaxy, and both galaxies together, we derive best-fit coefficients, $\alpha$, for linear combinations of \halpha{} and IR. For $12\mu$m, $70\mu$m, and $100\mu$m, these are one of only a few such estimates in the literature. For $8\mu$m, 24$\mu$m, and $70\mu$m, our results agree well with previous work, using a large set of individual regions. We also calculate the scatter around each model and tabulate those results in table \ref{tab:model STDS}.

Total {\sc Hi}+H$_2$ gas column density correlates well with \Ahalpha{}. $N(H)$ shows just as strong of a relationship with \Ahalpha{} as IR-to-\halpha{} color in these two galaxies. We compare our measurements to simple models relating gas column to \Ahalpha{}. Excluding galaxy centers, a screen with $\gamma=0.2$ times the total expected Milky Way attenuation describes data outside the galaxy centers reasonably well. This could be explained by $\sim 50\%$ of the dust lying behind the emitting regions and a $\sim 40\%$ spatial overlap between gas and \halpha{} emitting regions. Such a low overlap is consistent with recent results showing that \halpha{} and CO resolve into discrete distributions when observed at high physical resolution ($\sim 100$~pc). Direct, high resolution comparison of \Ahalpha{} to gas maps, e.g., with ALMA and JWST, should help further explore this issue.

\acknowledgements We thank the anonymous referee for a timely and constructive report that improved this work. We than Knox Long and Rick Pogge for sharing \halpha{} images of NGC~6946 and the HST helpdesk for helpful communications during the processing of the data. The work of SK, was supported by a grant from Program number HST-GO-14156. Support for Program number HST-GO-14156 was provided by NASA through a grant from the Space Telescope Science Institute, which is operated by the Association of Universities for Research in Astronomy, Incorporated, under NASA contract NAS 5-26555. This work was based on observations made with the NASA/ESA Hubble Space Telescope, obtained from the data archive at the Space Telescope Science Institute. STScI is operated by the Association of Universities for Research in Astronomy, Inc. under NASA contract NAS 5-26555. This paper almost makes use of project VLA/14A-171. The National Radio Astronomy Observatory is a facility of the National Science Foundation operated under cooperative agreement by Associated Universities, Inc. The work of SK and AKL is also partially supported by NASA ADAP grants NNX16AF48G and NNX17AF39G and the National Science Foundation grants No.~1615105, 1615109, and 1653300. This research made use of Montage. It is funded by the National Science Foundation under Grant Number ACI-1440620, and was previously funded by the National Aeronautics and Space Administration's Earth Science Technology Office, Computation Technologies Project, under Cooperative Agreement Number NCC5-626 between NASA and the California Institute of Technology.

\bibliography{sjk}

\appendix

\section{Details of the HST data processing}
\label{sec:HST data reduction}

This appendix details our processing of the HST data. To produce our final \pabeta{} line images, we processed the data using the HST pipeline, used 2MASS to calibrate the zero point of the images, checked the astrometric alignment of the images, estimated and subtracted the stellar continuum, masked foreground stars, and then mosaicked the individual panels into a single image for each galaxy. Much of this processing treated individual images (``panels''). We tabulate the central R.A. and Dec. of each panel in Table \ref{tab:panels}. 

\subsection{Correcting for non-linear backgrounds in the NGC~6946 data}
\label{sec:NGC 6946 sky correction}

For NGC~5194 we used standard MAST pipeline products. For NGC~6946, we inspected the FITS images produced by the pipeline. Many images in both filters showed large gradients across the chip. Correspondence with the WFC3 team at STScI revealed this to be caused by a time variable background (TVB). 

To correct for this, we re-ran the HST image alignment (\code{tweakreg}) and drizzling (\code{astrodrizzle}) pipeline with the original calibrated exposures, or ``flt'' images. In the ON filter, each calibrated exposure was inspected for TVB effects. Those that did show TVB issues were not included in the subsequent \code{tweakreg} and \code{astrodrizzle} pipelines. For the four fields, this led us to use only three of the four observed and calibrated exposures (panels 1, 3, 5, and 7 of the ON data). All of the OFF calibrated exposures were used because there were only two calibrated exposures per field. 

Once the TVB calibrated exposures were excluded, we used \code{tweakreg} to align the calibrated exposures before running the \code{astrodrizzle} pipeline. We found that the default sky subtraction method ('localmin') produced an all negative background. After trying each sky-subtraction method ('localmin', 'match', 'globalmin', and 'globalmin + match') we found the 'match' sky-subtraction method to yield the best result. This method, which computes differences in sky values between images in common sky regions and ``equalizes" sky values, was the only method that produced images absent of all negative backgrounds. We reran the \code{astrodrizzle} pipeline with these parameters. 

The resulting images were mostly free of noticeable TVB issues and negative backgrounds. However, visible issues still affected parts of three panels. These showed horizontal gradients which caused one half of the image to be brighter than the other. We excluded the affected parts of the image that from further analysis. These appear as missing regions in the figures in the main text (e.g., Figures \ref{fig:NGC5194 Mosaic} and \ref{fig:Aha_spatial}).

\subsection{Background matching and background subtraction}
\label{sec:onv2mass}

\begin{figure*}
    \centering
    \plottwo{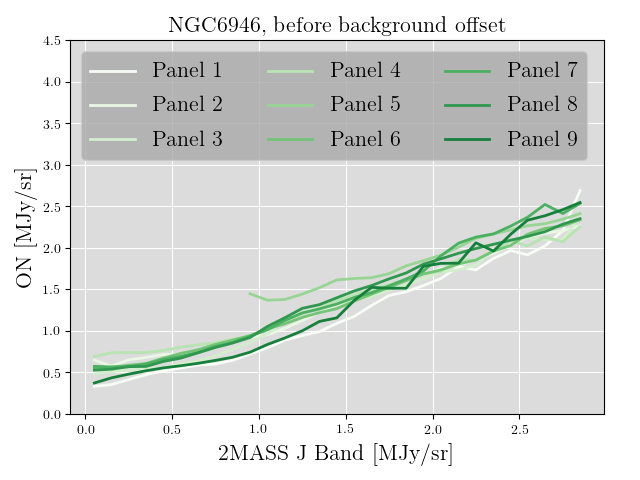}{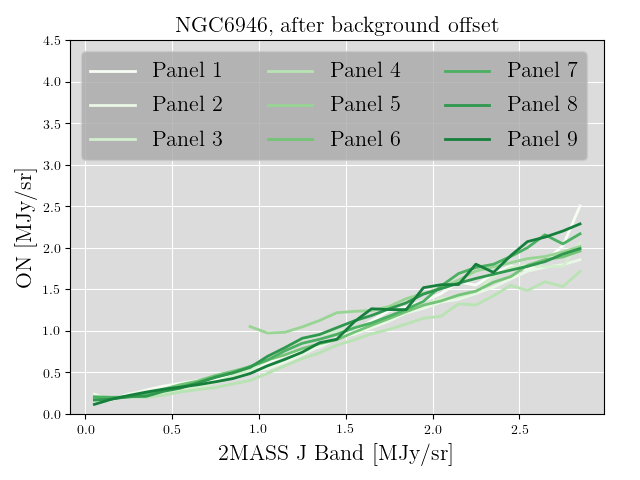}
    \plottwo{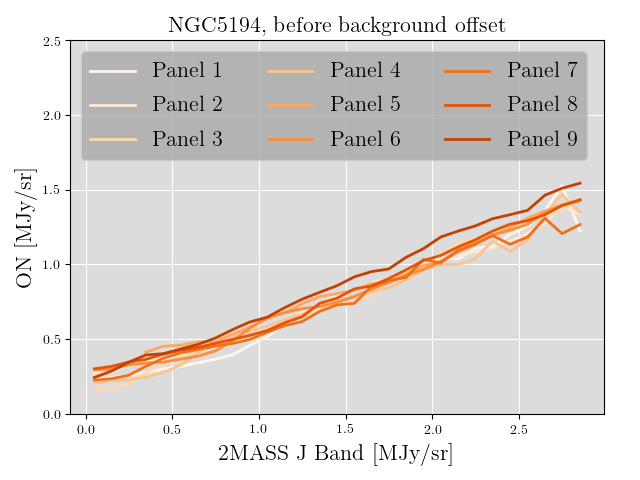}{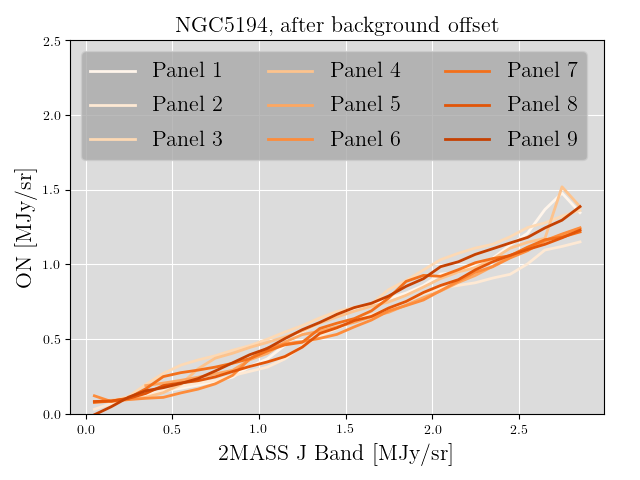}
    \caption{\textbf{Intensity of HST ON images as a function of 2MASS J band intensity before (top) and after (bottom) matching backgrounds.} We anchor the zero point of our HST ON band data to the wide field 2MASS J band images. After matching the resolution and astrometry of the data, we measure the average HST narrow band (ON, F128N) intensity in bins of 2MASS J band intensity. We calculate the offset needed to match the background of the HST images to that of 2MASS and subtract this from the ON images. The bottom row shows the result after this background matching.}
    \label{fig:onv2mass}
\end{figure*}

The field of view of HST is small compared to the size of our galaxies. This leads to a lack of empty sky in most frames, rendering the zero point of our images somewhat uncertain. We take two extra steps to establish a correct zero point and to ensure that the zero point of the ON and OFF images agree.

We anchor our background to J band images from the Two Micron All-Sky Survey (2MASS) \citep{SKRUTSKIE06} created by  \citet{JARRETT2013}. These images have wide area and excellent calibration and sky subtraction. The 2MASS J band filter has peak wavelength of 1.235\micron{} and overlaps both our ON and OFF filters. We matched the background in the high S/N ON images to 2MASS and then used the ON to set the background level in the OFF. 

First, we convolved our nine ON images with a Gaussian 2D Kernel of FWHM = 2.5\arcsec{} (the same PSF as the 2MASS images) and converted all images to have units of MJy~sr$^{-1}$. Then, we reprojected our data onto the 2MASS WCS and astrometrically aligned our images with the 2MASS data. Each image was inspected by eye in order to ensure proper alignment. 

Next, we measured the mean intensity of the HST ON data in bins of 2MASS intensity. Fig. \ref{fig:onv2mass} shows this comparison. The plot shows a good overall match between our HST images and 2MASS for both galaxies. However, in both cases our ON image still has some positive intensity as the 2MASS image goes to zero. Despite small mismatches in the filter, we expect both images to reflect the same galaxy structure and so go to zero together. Therefore, we calculated the offset needed to bring the ON into agreement with 2MASS for each panel. The average offset was $0.36$~MJy~sr$^{-1}$ for NGC~6946 and $0.41$~MJy~sr$^{-1}$ for the NGC~5194. For each panel, we subtracted the calculated offset from the HST ON images in order to match our background to that from 2MASS. The right panels in Figure \ref{fig:onv2mass} show the adjusted ON data, which agree well with the 2MASS data.  

Next, we match the background in the OFF to the ON. We do this, e.g., rather than independently matching the OFF to 2MASS, because for narrow band imaging purposes, good relative calibration between the line and continuum is essential to a stable result. To match the images, we convolved both the OFF and ON images to 1\arcsec. We chose this resolution because it is close to our final working resolution of $2''$, and by matching the images at this lower resolution we improve the signal to noise of individual pixels and ensure that any astrometric offsets we encounter (and fix) are relevant to the final analysis at $2''$ resolution.

\begin{figure*}\label{fig:background matching}
    \centering
    \plottwo{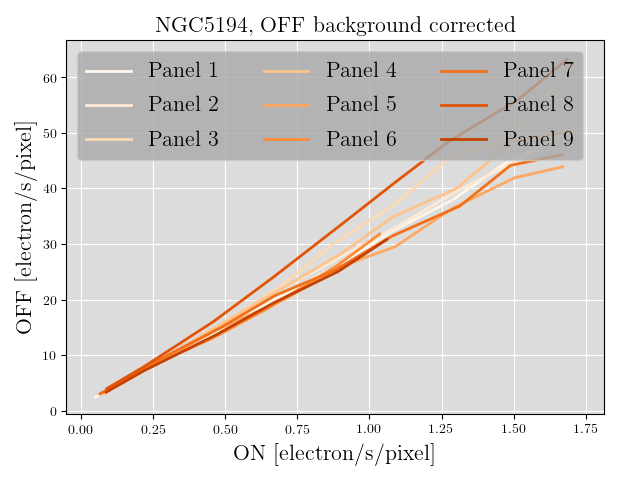}{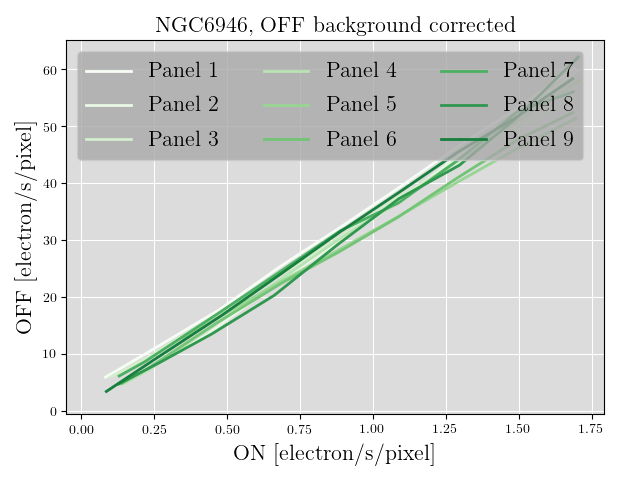}
    \caption{\textbf{Background matching between the HST ON and OFF images.} As Figure \ref{fig:onv2mass} but now only showing the OFF intensity as a function of on intensity after matching the zero points between the two images. }
    \label{fig:onvoff}
\end{figure*}

At this stage, we also inspected each image by eye to ensure proper alignment. Then, we matched the OFF to the ON using the same procedure that we used to match the ON to 2MASS. We compare the ON and OFF images after this match in Figure \ref{fig:onvoff}, this results in a good match between the ON and OFF data even down to low intensities. 

\begin{deluxetable}{lccccccccc}
\label{tab:panels}
\tabletypesize{\scriptsize}
\tablecaption{Panel centers}
\tablewidth{\linewidth}
\tablehead{
\colhead{Panel} &
\colhead{1} &
\colhead{2} &
\colhead{3} &
\colhead{4} &
\colhead{5} &
\colhead{6} &
\colhead{7} &
\colhead{8} &
\colhead{9}
}
\startdata
NGC~5194 & \\
... R.A. & 
202.4191 &
202.4265 &
202.4367 &
202.4572 &
202.4701 &
202.4784 &
202.4998 &
202.5098 & 
202.5181 \\
... Decl. & 
47.1564 &
47.1865 &
47.2193 &
47.1553 &
47.1878 &
47.2217 &
47.1570 &
47.1902 &
47.2210 \\
NGC~6946 & \\
... R.A. &
308.7449 &
308.6855 & 
308.6260 & 
308.6582 & 
308.7178 & 
308.7774 & 
308.8096 & 
308.7499 & 
308.6903 \\
... Decl. & 
60.1092 &
60.1275 &
60.1455 &
60.1718 &
60.1356 &
60.1356 &
60.1619 &
60.1799 &
60.1980 
\enddata
\end{deluxetable}


\subsection{Continuum subtraction} 
\label{sec: continuum subtraction}

\begin{figure*}
    \centering
    \includegraphics[width = 0.45\textwidth]{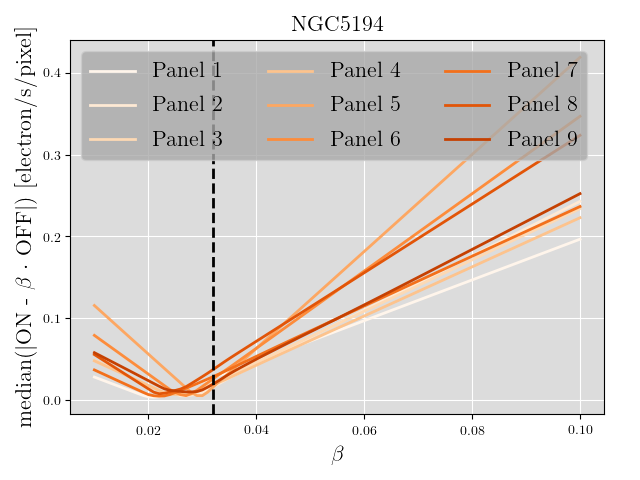}
    \includegraphics[width = 0.45\textwidth]{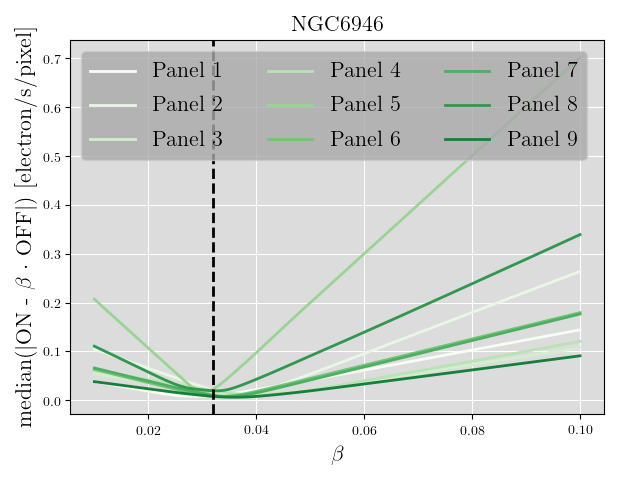}
    \caption{\textbf{Calibration of the scaling factor used to estimate stellar contamination in the ON using the OFF.} The $y$-axis shows the median absolute difference between the ON and scaled OFF ($\beta \cdot $OFF) for possible values of $\beta$ ($x$-axis). For each panel, we adopt the $\beta$ that yields the minimum median absolute value of the difference in regions picked to lack bright \halpha{}. We use this subtract the stellar continuum contribution to the ON image, yielding a final line-only image (Eq. \ref{eq:generalonoff}). The theoretical scaling factor (0.032) calculated by the ratio of the filter widths is plotted as a vertical black dashed line.}
    \label{fig:Ratios}
\end{figure*}

The ON images contain both \pabeta{} and stellar continuum emission. To isolate the \pabeta{}, we use the OFF image to estimate and subtract the contribution of stellar continuum to the ON. Emission in the OFF image represents almost all stellar continuum due to the much wider filter compared to the ON. Because the ON and OFF are so close in wavelength, and because we lack any additional constraints, we adopt a simple model in which the stellar continuum in the ON image is a scaled version of the OFF image. Then:

\begin{equation}
\label{eq:generalonoff}
I_{Pa\beta} = I_{ON} - \beta \cdot I_{OFF}~{\rm where}~    \beta = \frac{I_{ON,starlight}}{I_{OFF,starlight}}.
\end{equation}

We derive the scaling factor $\beta$ empirically, by comparing intensities between the ON and OFF image in regions where we expect both to contain only starlight. To isolate regions with only starlight, we exclude all lines of sight with bright \halpha{} (defined by $I(\halpha)>1$\e{-5}~erg~s$^{-1}$~cm$^{-2}$~sr$^{-1}$) from the analysis. Then, we create an array of possible $\beta$ and choose the one that minimizes the median absolute value of the difference between the ON and the scaled OFF. That is, we minimized the median $\mid ON - \beta \cdot OFF \mid$. 

We carried out this minimization for each panel. We show the results in Figure \ref{fig:Ratios}. We find average $\beta = 0.031$ in NGC~5194, with a $0.001$ rms panel-to-panel scatter. We find median $\beta = 0.034$ in NGC~6946, with rms scatter of 0.002. Then, we used the best-fit value of $\beta$ for each panel to subtract a scaled version of the OFF image to create a \pabeta{} line image. We inspected the final images to verify that our adopted $\beta$ did not over or under-subtract the continuum.

The \pabeta{} line contributes to the OFF image, but only weakly, $\lesssim 3.6\%$ on average based on our initial estimate of the line maps. This implies that we might slightly over-estimate the stellar continuum in regions where there is bright \pabeta{}. To correct for this, we scale our \pabeta{} image to be 3.6\% brighter than our initial estimate. 

We use a single, wide-band OFF. Ideally, we would estimate stellar contamination using multiple filters or a narrow band OFF close to ON in wavelength. Our filters lie in the J band near the Rayleigh-Jeans tail of stellar emission, so we do not expect major variations in the color of stellar populations. Modeling of dust attenuation would also benefit from data in multiple bands, but as \pabeta{} emits in the NIR ($\lambda = 1.284\micron$) we do not expect it to be heavily attenuated by dust. Similarly, we adopt a single $\beta$ per panel, while in reality we might expect the colors of the stellar population or dust structure to vary across a panel. Finally, we do not have a way to directly account for \textbf{any significant} \pabeta{} absorption features in the stellar continuum. We rely on our empirical calibration of the ON-to-OFF ratio to deal with this effect. All of these factors contribute at some level to the overall uncertainty in our final line image.

\subsection{Mosaicking, star masking, and convolution}
\label{sec: making pabeta}

After continuum subtraction, we reprojected the individual panels for each galaxy to create a single mosaic image. We used \code{Montage} to create the mosaic images of both NGC5914 and NGC~6946. We trust our earlier background subtraction and therefore we do not allow \code{montage} to change the background at this stage.
 
Both galaxies had foreground and saturated stars in the mosaicked images. As NGC~6946 lies at low Galactic latitude ($b= 11.7^\circ$) it so shows many more foreground stars than NGC~5194. We masked foreground and saturated stars by eye before any subsequent analysis or convolution was done.
 
After the foreground and saturated stars were masked, the moasicked images for both galaxies were convolved to a Gaussian 2D PSF of 2\arcsec. This matches the working resolution of our \halpha{} maps and ensures that the data used in our analysis have matched PSFs.
 
\subsection{Correction for Galactic extinction}
\label{sec:galacticav}

We correct both the \pabeta\ and archival \halpha\ for the effects of Galactic extinction. We adopt foreground extinction estimates from \citet{SCHLAFLY11}, via the NASA Extragalactic Database. Then we translate these from $A_V$ to \Ahalpha\ and \Apabeta\ using our adopted extinction curve. We scale the whole map for each line by this value. For NGC~5194 our best-estimate Milky Way \Ahalpha\ is 0.084~mag and our best-estimate for \Apabeta\ is 0.026~mag. For NGC~6946 the values are higher. We calculate Milky Way \Ahalpha\ of 0.810~mag and \Apabeta\ of 0.253~mag.

\subsection{Uncertainties in the final line maps}
\label{sec:noise}

\begin{figure*}
    \centering
    \includegraphics[width = 0.45\linewidth]{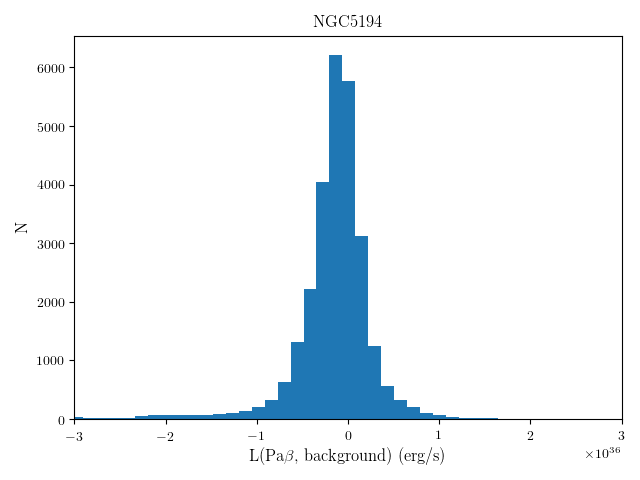}
    \includegraphics[width = 0.45\linewidth]{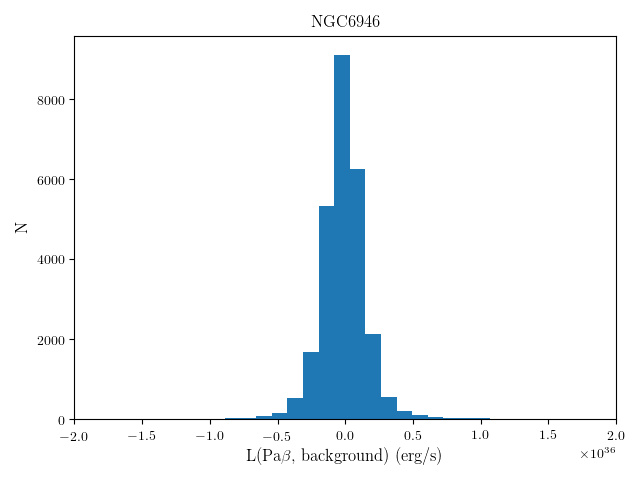}
    \caption{\textbf{Statistical noise estimates from NGC~5194 and NGC~6946 \pabeta{} maps.} Histograms of the luminosities of apertures placed in apparently empty regions of NGC~5194 (left) and NGC~6946 (right).}
    \label{fig:Background Hist}
\end{figure*}

We estimate the statistical noise in our final mosaicked images by measuring the standard deviation within an empty patch of sky at 2\arcsec{} resolution. In NGC~5194 the rms statistical noise is 1.46\e{-6} \intunits. In NGC~6946, the uncertainty is 5.63\e{-7} \intunits . In addition to these statistical uncertainties, the broad band filters (our OFF images) of HST's WFC3 have a reported calibration error of $\sim$2-3\% and the narrow band filters (our ON) have a reported photometric error of $\sim$5-7\%. Together, these imply an overall flux calibration uncertainty of $\sim 8-9\%$.
 
There is also an uncertainty associated with our continuum subtraction. The ON/OFF ratio used for subtraction can reasonably vary  by 5\e{-4}. To quantify this uncertainty we make six maps that vary from visibly over-subtracted to visibly under-subtracted. These maps are then used in our Monte Carlo analysis of our errors in Section \ref{sec:uncertainties}.

\end{document}